\documentclass[aps,prd,twocolumn,amsmath,groupedaddress,nofootinbib]{revtex4}

\usepackage{graphicx}
\usepackage{amssymb}
\usepackage{bm}
\usepackage{dcolumn}

\usepackage{color}   

\newcommand{\be}{\begin{eqnarray}}      \newcommand{\ee}{\end{eqnarray}}   
\newcommand{\ba}{\begin{array}}         \newcommand{\ea}{\end{array}} 
\newcommand{\bs}{\begin{subequations}}  \newcommand{\es}{\end{subequations}} 
\newcommand{\rf}[1]{~(\ref{#1})}  
\newcommand{\rfs}[2]{~(\ref{#1})-(\ref{#2})}

\newcommand{\ct}[1]{~\cite{#1}}  
\newcommand{\lb}[1]{\label{#1}}  

\newcommand{\nn}{\nonumber}
\newcommand{\lf}{\left}     \newcommand{\rt}{\right}
\newcommand{\fr}{\frac}     \newcommand{\ov}{\over}     

\newcommand{\la}{\langle}    \newcommand{\ra}{\rangle}    

\newcommand{\dst}{\displaystyle}

\newcommand{\Vsp}{\vphantom{\displaystyle{\hat I \over \hat I}}}
\newcommand{\vsp}{\vphantom{\displaystyle{I \over I}}}

\def\d{\delta}    \def\pd{\partial}  \def\D{\Delta}

\def\t{\tau}           
\def\al{\alpha}         
\def\g{\gamma}
\def\l{\lambda} 
\def\Nb{\nabla}

\def\rtg{\sqrt{-g}\,}
      
\def\f{\varphi}         

         \def\chf{\chi}
\def\Om{\Omega}        
\def\eps{\epsilon} 
\def\H{{\cal H}}
\def\A{{\cal A}}

\def\x{{\bm x}}
\def\^{\hat}            \def\~{\tilde}  
\def\n{{\bf n}}
\def\dOmn{d^2n_i}     

\def\eq{_{\rm eq}}     
\def\gdec{_{\rm \gamma\,dec}}      

\def\Rnu{R_{\nu}}           
\def\snu{_{\nu}}             

\def\eg{{\it e.g.\/}}   \def\cf{{\it c.f.\/}\ }   \def\ie{{\it i.e.\/}}

\def\i{_{\rm in}}
\def\gsp{{}^{(3)}\!g}
\def\di{\mathcal{I}}

\def\v{v}

\newcommand{\const}{{\rm const}}
\newcommand{\diag}{\mathop{\rm diag}}

\newcommand{\ci}{\mathop{\rm ci}}    \newcommand{\si}{\mathop{\rm si}} 
\newcommand{\Ci}{\mathop{\rm Ci}}    \newcommand{\Si}{\mathop{\rm Si}} 
    \newcommand{\Ip}{\mathop{\rm Im}}

\begin{document}

\title{\Large  Coupled Evolution of 
               Primordial Gravity Waves\\ 
               and Relic Neutrinos}

\author{Sergei Bashinsky}

\affiliation{International Centre for Theoretical Physics, 
          Strada Costiera 11, Trieste, Italy}

\date{\today}

\begin{abstract}
We describe analytically the cosmological evolution of primordial 
gravity waves (tensor perturbations) 
taking into account their coupling to relic neutrinos.  
We prove that as a consequence of causality
the neutrino-induced phase shift of subhorizon tensor oscillations
tends on small scales to zero.
For the tensor modes that reenter the horizon in the 
radiation era after neutrino decoupling
we calculate the neutrino suppression factor as 
$1-5\rho_{\nu}/9\rho+O[(\rho_{\nu}/\rho)^2]$.
This result is consistent with the value obtained 
for three neutrino flavors 
by Weinberg and is in agreement with numerical Boltzmann 
evolution.  A minimal formula with the correct asymptotic form  
on small and large scales reproduces 
to about ten percent the evolution 
on all scales probed by the CMB.  A more accurate solution
(in terms of elementary functions) shows that the modes reentering the horizon in 
the radiation era are slightly enhanced
and the phase of their temporal oscillations is shifted
by subdominant nonrelativistic matter.
The phase shift grows logarithmically on subhorizon
scales until radiation-matter equality;
the accumulated shift scales for 
$k\gg k\eq$ as $\ln k/k$.
The modes reentering the horizon after equality
are, in turn, affected by the residual radiation density.
These modes follow the naive matter-era 
evolution which is advanced by a redshift and scale independent
increment of conformal time.  
In an appendix, we introduce a general relativistic 
measure of radiation intensity  
that in any gauge obeys a simple transport equation and 
for decoupled particles is conserved on superhorizon scales 
for arbitrary initial conditions in the full nonlinear theory. 
\end{abstract}

\maketitle

\section{Introduction and overview of results}

The primordial power of gravity waves, or tensor 
metric perturbations, 
is a direct probe of the inflation energy 
scale\ct{Starob79,RubSazhVer82,FabPol83,AbbottWise84},
unknown at present.
Observation of their B-mode signature\ct{Kamionk96_Bmobe,SelZald_Bmode}
in polarization of the cosmic microwave background (CMB) 
is a target of several developed experiments.
Gravity wave modes
are generically frozen while the Hubble scale
is smaller than the mode wavelengths.
Yet, a mode starts to evolve as soon as it
``enters'' the Hubble horizon.  
This evolution determines the pattern of the tensor signatures 
in the CMB spectra. The modes processed by the evolution  
are to be constrained by ground and space-based 
interferometers, resonant mass detectors, or timing of 
millisecond pulsars.

While numerical calculations are helpful in theoretical studies
and may be necessary for data analysis, exact analytical results 
provide a framework for our understanding of
physical phenomena.
Quantitative analytic description of gravity wave 
evolution in the realistic cosmological setting,
roughly established today, is the subject of 
this paper.  A consistent, involving no fitting and ad hoc 
approximations, study in elementary formulas with
clear interpretation can be performed to 
several percent accuracy.

We invoke both the Fourier and real space views of 
cosmological dynamics.  
The advantage of Fourier decomposition 
is decoupling of modes with different wavenumbers
in linear order.
There are also merits in the 
complimentary real space approach.
First, the real space formulation
manifests the locality of
interactions and causality of solutions.
Causality imposes severe restrictions on 
the admissible evolution,  
and real space is a preferred starting ground to explore 
the implications.
Second, only in real space we have an appealing 
formulation of the full, non-linear, cosmological dynamics.
Understanding of the linear regime in real space 
is thus essential for connecting linear and non-linear theories.

Evolution of a tensor (transverse traceless, helicity $\pm2$) 
mode~$h_{ij}$ of the metric\rf{metric} 
with a wavevector~$k$ obeys a wave equation,
\eg \ct{Bardeen80,KS84,Bert93},
\be
\ddot h_{ij} + 2\H\dot h_{ij} + k^2 h_{ij} 
    = 16\pi Ga^2\Sigma_{ij}.
\lb{ddoth_i}
\ee
The mode dynamics is affected by the Hubble expansion rate  
$\H\equiv \dot a/a$ and the tensor  
component of matter anisotropic 
stress~$\Sigma_{ij}$\rf{Sigma_def}. 
After the horizon entry the mode undergoes 
continuous oscillations, which are damped by the Hubble expansion.  
On subhorizon scales ($k\gg \H$), 
the damping becomes adiabatic, $h\propto 1/a$, and 
the oscillation period settles at $2\pi/k$.  
The asymptotic amplitude and phase of the subhorizon
oscillations depend on the mode evolution during the horizon 
entry.

The modes with $k\gtrsim k\eq\equiv \t\eq^{-1}\sim 10^{-2}$\,Mpc$^{-1}$
enter the horizon when photons and neutrinos
make up a substantial fraction of the overall energy
density. This applies to all the modes accessible to direct detection
and to the modes exciting the CMB multipoles with 
$l\gtrsim d\gdec/\t\eq \approx 130$.
These modes also contribute appreciably to the
lower multipoles of CMB polarization
up to the rise of the reionization bump
at $l\sim 10$.

If anysotropic stress were negligible
then during radiation domination
$h^{\rm(rad,0)}=j_0(\f)=\sin \f/\f$.
Here and for the entire paper
\be
\f\equiv k\t
\ee 
and mode normalization is $h\to 1$ as $a\to0$.
In real space, the neutrinoless radiation era evolution of
gravity waves is described by
a top-hat Green's function
$h^{\rm(rad,0)}=\fr1{2\t}\,\theta(\t-|x|)$.

However, recently Weinberg\ct{Weinb_nu}  showed 
that the contribution of free-streaming neutrinos to
anisotropic stress and the consequent gravity wave 
damping\ct{Hawking66,KS84} is by no means tiny. 
The subhorizon amplitude of tensors and the induced 
by them CMB polarization were found suppressed by a factor 
$A_0\approx 0.8$\ct{Weinb_nu} on all scales entering the horizon
from neutrino decoupling to matter domination,
$\H_{\nu\,\rm dec}^{-1}\sim 0.1$\,kpc to $k\eq^{-1}$. 
All tensor power spectra are suppressed on these
scales by $A_0^2$.\footnote{
 Photon anisotropic stress, which develops after the recombination,
additionally suppresses a narrow range of tensor 
modes with $k\sim 0.005\,{\rm Mpc}^{-1}$ 
by up to $7\%$.  The affected are the modes that
enter the horizon after photons decouple but before   
the dynamical role of radiation becomes entirely negligible.
This effect, however, has little consequence for 
CMB polarization, generated earlier during the decoupling,
and even for the tensor ISW impact on the CMB temperature:
$C_l^{BB}$ and the tensor component of $C_l^{TT}$ are 
reduced by less than $0.5\%$ and than $1\%$
respectively.}

Weinberg's result was derived by calculating the 
anisotropic stress~$\Sigma$ of free streaming neutrinos as 
a line of sight integral over the metric perturbation~$h$.
Hence, eq.\rf{ddoth_i} was converted into a closed integro-differential 
equation.  The source on its right hand side, however,
was a complicated integral, which depended on $h_{ij}(\t)$ nonlocally 
and should be fully computed numerically at every evolution step. 
The amplitude and phase of the subhorizon 
oscillations could then be found as the ultimate output of numerical
evolution with this source from superhorizon to
subhorizon scales.

A real space approach provides an instant proof 
that regardless of the abundance of neutrinos, or presence
of additional exotic species, the phase
of subhorizon tensor oscillations in the radiation era 
is {\it unshifted\/} with respect to the zero-stress solution
$h^{\rm(rad,0)}$:
\be
h^{\rm(rad)}\to A_0\sin \f/\f
\lb{rad_asym_i}
\ee 
as $\f\to \infty$.\footnote{
  An analogous result exists in the scalar sector\ct{BS}.
On small scales, the phase of photon-baryon acoustic oscillations 
can be shifted only by 
gravitational coupling to species whose
perturbations propagate faster than the 
photon-baryon sound speed.  Among such species
are the free-streaming
neutrinos and a hypothetical scalar field with 
non-negligible density during radiation era (early quintessence).}
By causality of tensor metric perturbations\footnote{
   Propagation of gauge-dependent  {\it scalar\/} and 
{\it vector\/} perturbations of the metric, 
density, or a field generally 
is not causal, even if there is no physical violation of causality.
Nevertheless, when the initial conditions are {\it adiabatic},
scalar perturbations expand manifestly
causally in most conventional gauges, 
including the Newtonian gauge\ct{proof_adiab_causal}.
},  
their Green's function $h(\t,x)$ 
which is localized at the origin at $\t\to0$ 
should remain zero beyond the particle horizon:
\be
h(\t,x)=0\quad \mbox{for}\quad |x|>\t.
\lb{causal_i}
\ee
(We imply that tensor perturbations develop
entirely from primordial metric inhomogeneities, \ie,
there are no primordial ``isocurvature'' tensor excitations.) 
Subhorizon oscillations of the Fourier modes 
are mapped to small-scale discontinuous features
of the Green's function at the horizon $|x|=\t$.  
Only when the phase shift 
is absent [$\f_0$ of eq.\rf{asymp_mode_gen} is zero] 
the corresponding, step-like, discontinuity [eq.\rf{h_asym}] 
is consistent with condition\rf{causal_i}.

The amplitude suppression factor~$A_0$ 
is determined by the Green's function jump at the particle horizon.
In the order linear in $\Rnu\equiv \rho\snu/\rho$,
when neutrino stress is calculated ignoring
its feedback to the metric,
we find
\be
A_0 = 1-\fr59\,\Rnu.
\lb{A0_i}
\ee
For three neutrino flavors 
($\Rnu\simeq 0.405$ for 
$N_{\nu\,\rm eff}\simeq 3.04$\ct{Dicus_NuID,Heckler_NuQED,LopezTurner_NuQED})
the agreement of eq.\rf{A0_i} with a numerical
value $A_0^{\rm (CMBFAST)}\simeq 0.76$ is $2\%$.

Fourier transformation of the Green's function,
which we derive in the $O(\Rnu)$ order,
gives an analytic expression for the evolution of tensor
modes in the radiation era.
Good match of corresponding
equations~(\ref{h0_rad},\ref{h1k})
with the full numerical {\sc cmbfast}\ct{cmbfast}
calculations is demonstrated on the right panel of Fig.~\ref{fig_rad}.  

The amplitude and phase of gravity wave oscillations
are affected by changes in the Hubble expansion
during the subsequent cosmological epochs.
On the smallest scales ($k/\H\to\infty$)
the amplitude and phase shift evolve adiabatically
(infinitesimally over an oscillation period).
Then the amplitude decays as $1/a$ 
while the phase shift remains frozen. Hence,
\be
h(\t,k\to\infty)= A(\t)\sin \f/\f
 =A(\t)\,h^{\rm(rad,0)}(\f),
\lb{large_k_i}
\ee
where $A(\t)$ is given by eq.\rf{A_t}.
In real space, this small-scale behavior 
corresponds to a step discontinuity
of Green's function at the particle horizon,
\be
h^{\rm (disc)}(\t,x)={A(\t)\ov 2\t}\,\theta(\t-|x|)
                 =A(\t)\,h^{\rm(rad,0)}(\t,x).
\lb{disc_i}
\ee

The modes entering the horizon in the matter era
($k\ll k\eq$) encounter negligible anisotropic
stress.  For them during matter domination
\be
h^{\rm(mat,0)}=3j_1(\f)/\f
              =3\lf(\,\sin \f/{\f^3}-{\cos \f}/{\f^2}\rt).
\lb{h_mat_i}
\ee
The respective Green's function is 
\be
h^{\rm (mat,0)}={3\ov 4\t}\lf(1-{x^2\ov\t^2}\rt)\theta(\t-|x|).
\lb{grf_mat_i}
\ee

Real space results\rf{disc_i} and\rf{grf_mat_i} 
suggest a compact analytic expression
that correctly describes the tensor evolution in
both the small and large scale limits and is qualitatively
acceptable in the intermediate range:
\be
h\sim  A(\t)\,h^{\rm(rad,0)}+[1-A(\t)]\,h^{\rm(mat,0)}.
\lb{h_min_i}
\ee
At any $\t$, the first term guarantees the proper $k\to\infty$
behavior\rf{large_k_i}, while on superhorizon scales $\mbox{$h(k\to0)=1$}$.
(The real space equivalent of the latter condition is $\int h(x)\,dx=1$.) 
In the matter era,
in agreement with further accurate analysis,  
the phase shift of subhorizon modes 
varies with~$k$ but is constant in time, 
eq.\rf{hmin_mat_subhor}.  
Expression\rf{h_min_i} reduces to $h^{\rm(mat,0)}$ for a mode 
that enters the horizon in the matter era, when $A(\t)$ decays.
This simple parameterization
matches well the numerical tensor evolution to the 
CMB last scattering and beyond,
Figs.~\ref{fig_glob} and~\ref{fig_glob2}.
Yet, despite the close match,  
formula\rf{h_min_i} is only qualitative for a finite~$k$.
A systematic treatment, presented further, reveals important 
effects eluding this minimal description.

As an alternative to eq.\rf{h_min_i}, a very popular parameterization 
of tensor evolution by Turner, White and Lidsey\ct{TWL93} 
considers an ansatz $h^{\rm TWL}= T(k/k\eq)\,h^{\rm(mat,0)}$.
Here, a ``transfer function'' is  $T(y)=[1+1.34y+2.5y^2]^{1/2}$,
with its last two coefficients fixed by a numerical fit
at $z=0$.  
This parameterization is rather misleading.  
Most importantly,
the ansatz $T(k/k\eq)\,h^{\rm(mat,0)}$ is motivated
by an erroneous statement 
that ``for $\t\gg\t\eq$, the temporal behavior of {\it all\/} modes 
is given by $3j_1(k\t)/k\t$''.
On the contrary, both subhorizon solutions of the
wave equation in the matter era, 
$h^{\rm(mat,0)}\to -3\cos\f/\f^2$ and
the complimentary $n\to3\sin\f/\f^2$, eq.\rf{mat_sol_compl}, 
decay as $1/a$.
A mode with $k\gtrsim k_{\rm eq}$ for $\t\gg\t\eq$ 
is described by their linear combination, 
which is determined by the evolution before matter domination.  
In fact, only the complimentary solution
contributes when $k\gg k\eq$, eq.\rf{large_k_i}.
This ironic feature and the importance of the phase shift 
for the CMB signatures
of gravity waves was emphasized by Ng and Speliotopoulos\ct{NgSpel94}.

Other shortcomings of the TWL transfer function are
disregard of the neutrino-induced suppression of the modes with 
$k\gtrsim k\eq$ and of the dynamical effects caused by 
the residual radiation 
density around the CMB last scattering
($\rho_{\rm rad}/\rho\,|\gdec\approx 0.25$ 
for the WMAP parameters\ct{WMAPSpergel}). 
The second problem was alleviated by 
Wang's\ct{Wang95} time-dependent transfer functions
which contain two additional fitting parameters.
Nevertheless, his parameterizations continued 
to evolve to the incorrect $\cos\f/\f^2$ form
even for the shortest wavelengths.

We obtain the functional form of the tensor evolution at a finite~$k$
by analytically solving the dynamical equations under controlled 
approximations. In addition to leading naturally 
to adequate simple analytical description,
derivation, as opposed to fitting, also offers 
better means of identifying the physics responsible for 
evolution features.
For earlier works that applied different methods 
but were guided by the same intentions see 
Refs.\ct{NgSpel94,KorAll94_h_analyt,PritchKam04}.

The modes that enter the horizon before equality ($k\t\eq>1$)
offer a surprise.  Contrary to intuitive expectations, the phase 
shift of their subhorizon oscillations grows {\it after\/} the horizon entry
throughout the rest of the radiation epoch.
The growth is due to the increasing 
role of nonrelativistic matter in the Hubble expansion. 
When matter becomes dominant,  
the phase shift saturates at a value\rf{Df_sub_mat}.

We analyze the subhorizon evolution with the WKB expansion
in the orders of $k/\H$. 
To start with an order above the 
adiabatic approximation, we factor out the $1/a$ decay 
by considering a variable $\v\propto ah$,
eq.\rf{ddot_f_a}\ct{Grishch74}.
Under the horizon, the amplitude of $v$
oscillations freezes.  
However, their phase shift
grows during radiation domination 
as $\fr1{2k\t_e}\ln\f+\const$, 
where $\t_e=\t\eq/(\sqrt2 - 1)\sim  260$\,Mpc
is given by eq.\rf{t_e_def}.
The oscillation amplitude and the additive constant in the phase 
are set by the mode evolution during the horizon entry,
described next. 

Matching the next-to-adiabatic WKB order 
to the superhorizon primordial fluctuations
is rather subtle.
For initial oscillations the WKB approximation 
breaks down and $\v^{\rm WKB}$ diverges.
Ng and Speliotopoulos\ct{NgSpel94} approached the problem by
changing the evolution variable to $\ln\t$,
now varying from $-\infty$.
They matched the growing mode in the ``forbidden'' superhorizon
region to the ``allowed'' subhorizon one
by applying the standard WKB matching procedure at a turning point.  
Their result reproduced accurately the phase of subhorizon 
oscillations but significantly underestimated the amplitude.
Pritchard and Kamionkowski\ct{PritchKam04} stretched
the WKB approach by retaining the decaying mode
in the forbidden region.
The obtained in terms of special functions
amplitude remained in poor agreement with numerical
calculations.

We consider the WKB solution for $v(\t)$ only on
subhorizon scales.
To normalize its amplitude and fix the phase, 
during the horizon entry we, instead, treat the matter correction 
to the Hubble expansion perturbatively in $\rho_{\rm mat}/\rho$.
Both the WKB and perturbative approximations are valid 
and both solutions can be matched
when a mode has entered the horizon but the matter
density is still negligible.  
The result, in terms of elementary functions,
\be
h=A(\t)\sin[\f-\D\f(\t)]/\f
\ee
with the amplitude suppression\rf{A_sub_radent} and phase shift\rf{Df_sub}
is in excellent agreement
with {\sc cmbfast} computations (Fig.~\ref{fig_rad_ent}). 
As a comparison, the matching procedure of\ct{PritchKam04} reproduces 
the amplitude of subhorizon oscillations 
for $k=0.1677$\,Mpc$^{-1}$ to $13\%$\ct{PritchKam04}. 
For the same~$k$ and same cosmological 
parameters, our formula achieves~$1\%$.

The evolution of the modes that enter the horizon 
after equality ($k\t\eq<1$) is likewise affected 
by the subdominant radiation density.
The effects in this case are less dramatic.
In the linear in $k\tau\eq$ order the impact of the residual 
radiation is fully accounted for 
by a shift $\t\to\t+\t_e$ in the radiation-free transfer
function: $h(\t)\approx h^{\rm(mat,0)}(\t+\t_e)$. 
This result is obtained by observing that
for $\t\lesssim\t\eq$ the large-scale scale modes are essentially 
frozen and that for $\t\gg\t\eq$ the background expands as
$a(\t)\approx a^{\rm(mat,0)}(\t+\t_e)$. 
The advance of the tensor evolution on the large scales
by the constant conformal time increment~$\t_e$
is consistent with numerical computations~(Fig.~\ref{fig_mat_ent}).

Tensor metric perturbations were connected to the CMB temperature 
and polarization spectra in the original 
works\ct{SachsWolfe67,DoroshNovikPoln77,Starob84,Poln85_tens_polar}, 
a full systematic approach was developed in\ct{Zald_Selj_allsky97}, 
and a review of the physical 
picture and typical analytic approximations
given recently in\ct{PritchKam04}.
We do not elaborate this connection in the present article.
Yet, we remember that
generation of CMB polarization 
and its curl-type $B$~component,
clean from linear scalar 
contribution\ct{Kamionk96_Bmobe,SelZald_Bmode}, 
requires photon scattering on free electrons, 
to polarize photons, but moderate optical depth,
for the polarized photons to reach the Earth.  
These conditions are met inside
the surface of CMB last scattering ($z\sim 1090\pm 100$)
and after reionization ($z\lesssim 20-6$).
The tensors also contribute to CMB temperature
anisotropy, primarily, through the integrated Sachs-Wolfe
mechanism\ct{SachsWolfe67}, induced by $\dot h_{ij}$
after the last scattering ($z\lesssim 1090$).

Among other results of this paper, we note simple equation\rf{dotI_gen}
that describes fully general-relativistic transport of radiation 
of decoupled ultrarelativistic particles.
The radiation intensity is quantified by 
a variable~$I(x^\mu,n_i)$, 
uniquely specifying 
the particle energy-momentum tensor with eq.\rf{Tmunu_int}.  
An advantage of the suggested intensity definition~$I$,
given in Appendix~\ref{gen_free_stream},
is that $I$~transport is affected only by 
spatial gradients of the metric and not by
metric temporal change.
As a result, the intensity  and its statistical distribution
are time-independent on superhorizon scales
in any gauge.
This fully nonlinear conservation does not require 
adiabaticity of cosmological perturbations.
If particles have ever been in thermal equilibrium 
and decoupled on superhorizon scales,
a superhorizon value of their intensity~$I$ is simply related 
to metric perturbation on a hypersurface of uniform
particle density, eq.\rf{ic_gen_a}.
In the locally Minkowski metric $I$ reduces to the
regular particle intensity $dE/(dVd\Omega_{\^n})$.

The rest of the paper is organized as follows.
Sec.~\ref{sec_dyn} reviews the dynamical equations.
Sec.~\ref{sec_rad} solves the coupled evolution of 
gravity waves and neutrinos during the radiation era.
Sec.~\ref{sec_mat} considers the evolution
of tensor perturbations through the radiation-matter transition
to the matter era. 
In Appendix~\ref{gen_free_stream} we describe propagation 
of decoupled particles in an arbitrary metric
and in linear theory.
In Appendix~\ref{appx_proof} we confirm by explicit
integration of the real space dynamics that no amount
of free streaming neutrinos 
can change the gravity wave phase in the radiation era.

In this paper, space-time coordinates $x^\mu=(\t,\x)$
correspond to the perturbed Friedmann-Robertson-Walker metric
\be
g_{\mu\nu}=a^2(\t)\lf[\eta_{\mu\nu}+h_{\mu\nu}(x^\mu)\rt],
\lb{metric}
\ee
with $\eta_{\mu\nu}=\diag(-1,1,1,1)$.
Zero of conformal time~$\t$ is set
by the condition $\t(a\,{\to}\,0)\equiv 0$ in the 
radiation era.  The scale factor $a\equiv 1$ at present.  
Background curvature, if any, is ignored
for the considered redshifts.
Figures compare the
analytic formulas with numerical
computations assuming a flat model with 
$\Omega_{\rm m}=0.3$, $\Omega_bh^2=0.022$, and $h=0.7$.

\section{Dynamics}
\lb{sec_dyn}

The linear evolution of gravity waves  
is described by wave 
equation\rf{ddoth_i}.
The gravity waves are driven by neutrino anisotropic 
stress, which develops after neutrinos decouple
($z< z_{\nu\,\rm dec}\sim 10^{10}$).
The driving is
dynamically relevant while neutrinos contribute noticeably
to the total energy density ($z\gtrsim z\eq\approx 3200$). 
At those redshifts, 
all the three standard neutrino generations 
may be assumed ultrarelativistic,
given the current cosmological bounds on the 
neutrino masses\ct{WMAPSpergel,Hannestad03_nu,Allen03_mnu_signal,
TegmarkSDSS03,Barger03_mnu,Crotty04_mnu,SeljakSDSS04,Fogli04_mnu}.
Then neutrino dynamics is approachable 
in the full general relativity for an arbitrary metric,
as discussed in Appendix~\ref{gen_free_stream} and summarized 
by Sec.~\ref{ss_ful_transp} next.

\subsection{General relativistic free streaming}
\lb{ss_ful_transp}

We describe the transport of ultrarelativistic neutrinos
by intensity $I$, defined in Appendix~\ref{gen_free_stream} 
as their energy-integrated phase space distribution.
$I(x^{\mu},n_i)$ is a function of spacetime coordinates and
direction of neutrino propagation.
The direction is specified by the covariant spatial
components $n_i$ of a null vector $n^{\mu}\propto d x^{\mu}/d\t$
which is normalized as $\sum_{i=1}^3 n_i^2=1$.

The impact of neutrinos on the metric is determined 
by their energy-momentum tensor~$T^{\mu }_{\nu}$.
For a given intensity~$I$,
\be
T^{\mu }_{\nu}
   =\int\fr{\dOmn}{\rtg}\fr{n^\mu n_\nu}{n^0}\,I(x^{\mu},n_i),
\lb{Tmunu_int}
\ee
eq.\rf{Tmunu_int_a}, where $n_0(n_i)$ can be found from 
the null condition~$n_\mu n_\nu g^{\mu\nu}=0$.
Provided the considered perturbations are superhorizon 
during neutrino decoupling, 
the initial conditions are set on a hypersurface 
of uniform neutrino density by\rf{ic_gen_a}
\be
I^{(u)}(\x,n_i)=\fr{\const}{\lf[n_in_j\gsp^{ij}_{(u)}(\x)\rt]^2}.
\lb{ic_gen}
\ee
Afterward, the neutrino intensity evolves according to transport 
equation\rf{dotI_gen_a}
\be
n^\mu\fr{\pd I}{\pd x^\mu}= \fr12\,n^\mu n^\nu g_{\mu\nu,i}
     \lf(4n_iI-\fr{\pd I}{\pd n_i}\rt).
\lb{dotI_gen}
\ee

The considered intensity~$I$ is gauge dependent.\footnote{
  Similarly to the majority of the traditional cosmological 
  variables, \eg\ct{Bardeen80},
  $I(x^\mu,n_i)$ defines a ``gauge-independent'' variable
  in any fixed gauge. 
} 
However, by separating the 
particle and metric dynamics, the presented description  achieves, 
first, compact transport equation in the full theory.  
Second, time independence of the intensity for 
superhorizon inhomogeneities, with negligible 
$g_{\mu\nu,i}$. (The metric, which is partly 
determined by a gauge condition, may nevertheless evolve even 
on superhorizon scales.)
In a locally Minkowski metric, $I$ reduces to the conventional 
intensity of neutrino radiation. 

In linearized theory, the intensity perturbation $\di\equiv \d I/I$
obeys transport equation\rf{dotD_a}
\be
\dot \di + n_i\di_{,i}  
  = 2n_i\^n^\mu \^n^\nu h_{\mu\nu,i},
\lb{dotD}
\ee
with  $\^n^\mu \equiv (1,n_i)$ and initial conditions\rf{ic_lin_a},
\be
\di^{\!(u)}=2n_in_jh_{ij}^{(u)}.
\ee
The perturbed components of the neutrino energy-momentum tensor
are determined by $\di$ from eqs.\rfs{T00pert_a}{Sigma_gen_a}.

\subsection{Tensor modes}
\lb{ss_tens_dyn}

Tensor linear perturbations over the flat Friedmann-Robertson-Walker
metric satisfy the conditions
$h_{00}=h_{0i}=0$ and
\be
\pd_ih_{ij}=0,  \qquad 
\sum_ih_{ii}=0.
\lb{tens_cond}
\ee
The tensor perturbations evolve according to 
a wave equation, \eg \ct{Bardeen80,KS84,Bert93},
\be
\ddot h_{ij} + 2\H\dot h_{ij} - \Nb^2 h_{ij} 
    = 16\pi Ga^2\Sigma_{ij},
\lb{ddoth}
\ee
where they are sourced by the tensor (transverse) component 
of anisotropic stress
\be
\Sigma^i_j\equiv T^i_j - \fr13\,\d^i_j\, T^k_k.
\lb{Sigma_def}
\ee

Neutrino contribution to anisotropic stress
follows from eq.\rf{Sigma_gen_a} by substituting the explicit 
linear solution
for neutrino intensity\rf{D_los_a}. 
When $\t\i$ in this solution is set to zero and the above
tensor  conditions are applied, eq.\rf{D_los_a}
gives\ct{Weinb_nu}
\be
\fr12\,\di(\t,\x)= n_in_jh_{ij}(\t,\x)
        -\int_{0}^{\t}\! d\t'\, n_i n_j \dot h_{ij}(\t',\x'),
\lb{D_los_t}
\ee
with $\x'= \x+\n(\t'-\t)$.

The impact of photon anisotropic stress on tensor metric 
perturbations is not entirely negligible.
While local photon anisotropy is washed out by Thomson 
scattering before the recombination, the anisotropy 
develops after photons decouple at $z\gdec\sim 1090$.
At this redshift, radiation constitutes as much as a quarter 
of the total energy density.
The generated photon stress affects the modes that
enter the horizon shortly after the decoupling.
Numerical calculations show up to $7\%$ consequent suppression 
of the metric modes with $k\sim 0.005\,{\rm Mpc}^{-1}$.
However, since the photon stress was insignificant 
during decoupling, CMB polarization is almost unsuppressed.
{\sc cmbfast} predicts less than $0.5\%$ suppression of the
polarization correlation~$C_l^{BB}$.  The photon stress
also affects little CMB temperature -- the corresponding 
suppression of the tensor component of $C_l^{TT}$ is less than~$1\%$.
We therefore neglect photon anisotropic stress in the following
analysis.  Even more so, we can disregard the stress of
nonrelativistic species (baryons and CDM) 
because of their tiny velocity dispersion.

Now we consider a single plane wave,
not necessarily harmonic, and choose the coordinates $y$
and $z$ in a plane of constant $h_{ij}$:
$h_{ij}=h_{ij}(\t,x)$.
Then for tensor perturbations,
by the first of conditions\rf{tens_cond},
$h_{xi}=h_{ix}=0$ and
$\Sigma_{xi}=\Sigma_{ix}=0$. For the remaining
components of $\Sigma^i\snu{}_j$ with $i,j\in(y,z)$
we can take the average in eq.\rf{Sigma_gen_a} 
over the orientation of $(n_y,n_z)$ under a fixed $n_x\equiv \mu$.
The result is
\be
\Sigma^i_j(\t,x) = -\fr{\rho\snu}{4}\!\int_0^{\t}\!\! d\t'\!
               \int_{-1}^1\!\!\!d\mu\,(1-\mu^2)^2
               \dot h_{ij}(\t'\!,x'),~~
\lb{Sigma_t}
\ee
where $x'\equiv x+\mu(\t'-\t)$.
Fourier transformation of this expression
reproduces eq.~(16) of Ref.\ct{Weinb_nu}.

From now on, we imply a decomposition of a gravity wave $h_{ij}$ 
over some polarization basis, \eg, into $+$ and 
$\times$ components,
\be
h_{ij}=\sum_{\l=1,2}\eps_{ij}^{(\l)}h_{\l}.
\ee
Since the evolution imposed by eqs.\rf{ddoth} and\rf{Sigma_t} 
is identical for all the polarization components,
we consider an arbitrary component~$h_\l$ 
and drop the subscript~$\l$ later.

\subsection{Plane tensor Green's functions}
\lb{ss_grf}

The evolution of any linear perturbation can be described
by superposing plane Green's functions\ct{BB_PRL,BB} 
$h(\t,x)$, which satisfy the initial condition
\be
h(\t\to0,x)=\d(x).
\lb{Gf_ic}
\ee
For example, a harmonic plane wave mode
$h(\t,k)\,e^{ikx}$ with the initial condition $h(\t\to0,k)=1$ 
is described by the Fourier integral
\be
h(\t,k)=\int^{\infty}_{-\infty} dx\, e^{-ikx}\, h(\t,x).
\lb{sumrl_gen}
\ee

Since $h\wedge\dot h$ decays, or ``squeezes''\ct{Grishchuk_squeez90}, 
on superhorizon scales,
the initial $\dot h$ need not be specified precisely.
It is sufficient to require a finite $\t\to0$ limit of
$\int_{-\infty}^{\infty}dx\,\dot h(x)$,
to ensure that we work with a non-decaying mode.

The initial condition\rf{Gf_ic}
has several immediate consequences.
First, by the causality of gravity wave propagation
\be
h(\t,x)=0\quad \mbox{for}\quad |x|>\t.
\lb{causal}
\ee
Second, by parity conservation during the considered evolution,
the initially even perturbation\rf{Gf_ic}
remains even: $h(\t,-x)=h(\t,x)$.
Third, the $k\to0$ limit of eq.\rf{sumrl_gen},
time-independence of $h(\t,k\to0)$, and 
initial condition\rf{Gf_ic}
add up to a sum rule
\be
\int^{\infty}_{-\infty} dx\ h(\t,x)=h(\t,k\to0)=1~~~
\lb{sum_rule}
\ee
for all $\t$.

\section{Radiation era}
\lb{sec_rad}

The magnitude and phase of the tensor modes with 
$k\gg k\eq$ is set by
their dynamics in the radiation epoch.
At that time neutrino anisotropic stress plays an 
active role.  The coupled dynamics of 
gravity waves and neutrinos during radiation domination 
can be solved analytically in real space.
Analytic expressions for the radiation era
evolution of Fourier tensor modes 
then follow by Fourier transformation.

\subsection{Self-similarity}
\lb{ss_self_sim}

The coupled evolution of gravity waves and neutrinos
with the specified Green's function initial 
conditions involves no external dynamical scales
during radiation domination, when $\H=1/\t$.
Then the plane tensor Green's function,
whose dimension is $\t^{-1}$, takes a self-similar form 
\be
\qquad h(\t,x)=\fr{\bar h(\chi)}{\t},\qquad
\chi\equiv \fr{x}{\t}.
\lb{h_selfsim}
\ee
The initial condition\rf{Gf_ic}
requires the normalization
\be
\int_{-1}^{1}d\chi\, \bar h(\chi)=1,
\lb{ss_norm}
\ee
as evident from the sum rule\rf{sum_rule}.
Note that integration over $\chi$ is
restricted to $\chi\in[-1,1]$
because by causality\rf{causal} 
$\bar h(\chi)$ vanishes identically 
beyond this interval.

We determine
$\bar h(\chi)$ by solving eqs.\rf{ddoth} and\rf{Sigma_t}. 
We use that
$\dot h = - (\chf\bar h)'/\t^2$, 
$\ddot h = (\chf^2\bar h)''/\t^3$,
$\Nb^2  h = \bar h''/\t^3$,
where primes denote $\chf$~derivatives,
and  in the radiation era 
$8\pi Ga^2=3/(\rho\t^2)$ by the Friedmann equation.
With these substitutions, eqs.\rf{ddoth} and\rf{Sigma_t}
give:
\be
&&\lf[(\chi^2-1)\bar h'\rt]'\,  = 
\lb{sseq_1}
\\
&&=\,\fr{3\Rnu}2\int_0^{\t} \!\! \fr{\t d\t_1}{\t_1^2} 
     \int_{-1}^1d\mu\,(1-\mu^2)^2
     \fr{d[\chi_1\bar h(\chi_1)]}{d\chi_1}, 
\qquad
\nn
\ee
where $\chi_1\equiv (\chi-\mu)\fr{\t}{\t_1}+\mu$
and $\Rnu\equiv \rho\snu/\rho$.
{\it Carefully\/} integrating
over $d\t_1$ by parts and remembering the causality\rf{causal}
we obtain:
\be
&&\lf[(\chi^2-1)\bar h'\rt]' \,  = 
\lb{sseq}
\\
&&=\,\fr{3\Rnu}2\ \theta(1-|\chi|)\int_{-1}^1d\mu\,[K(\mu,\chi)-K(\chi,\mu)],
\qquad
\nn
\ee
where $\theta$ is the Heaviside step function and 
\be
K(\mu,\chi)\equiv \fr{(1-\mu^2)^2\chi \bar h(\chi)}{\mu-\chi}.
\lb{K_def}
\ee

Equation\rfs{sseq}{K_def} is singular
at $\chi=\pm1$ and $\chi=\mu$. 
The real-space singularities are treated consistently 
with the Fourier mode approach by the standard formalism
of generalized functions
(see discussion in Sec.~IV.C of Ref.\ct{BS}).  
In the following analysis of the tensor sector 
it is sufficient to remember that, first, 
the $d\mu$ integral in eq.\rf{sseq} should be taken in
the Cauchy principal value. 
Second, an equation  $(\chi-a)A=B$ for any
generalized functions $A(\chi)$ and $B(\chi)$ 
should be resolved as
$A={B}/(\chi-a)+\const\,\d(\chi-a)$,
where the last term is the Dirac delta function.
The delta function prefactor can be
determined from the properties
imposed on~$A$ 
by the initial conditions; for example, 
from the sum rule\rf{sum_rule}.

\subsection{Phase shift}
\lb{sec_phase}

Without neutrino anisotropic stress ($\Rnu=0$),
the even normalized solution of eq.\rf{sseq} is
\be
\bar h^{(0)}=\fr12\,\theta(1-|\chi|).
\lb{h0}
\ee
Fourier transformation of the 
corresponding Green's function\rf{h_selfsim} 
gives the standard neutrinoless radiation era result
\be
 h^{(0)}(k\t)=\fr{\sin(k\t)}{k\t}.
\lb{h0_rad}
\ee 

Although neutrinos in the radiation era
are, in fact, among the dominant species, 
the gravitational impact of their anisotropic stress becomes 
negligible on subhorizon scales.
The corresponding asymptotic ($k\t\gg1$)
homogeneous solution of 
mode evolution eq.\rf{ddoth_i} 
is
\be
 h(k\t)=\fr{A_0\sin(k\t+\f_0)}{k\t}+O\lf((k\t)^{-2}\rt).
\lb{asymp_mode_gen}
\ee 
We prove that, regardless of an $\Rnu$ value, $\f_0=0$.

\begin{figure*}[tb]
\includegraphics[width=15.5cm]{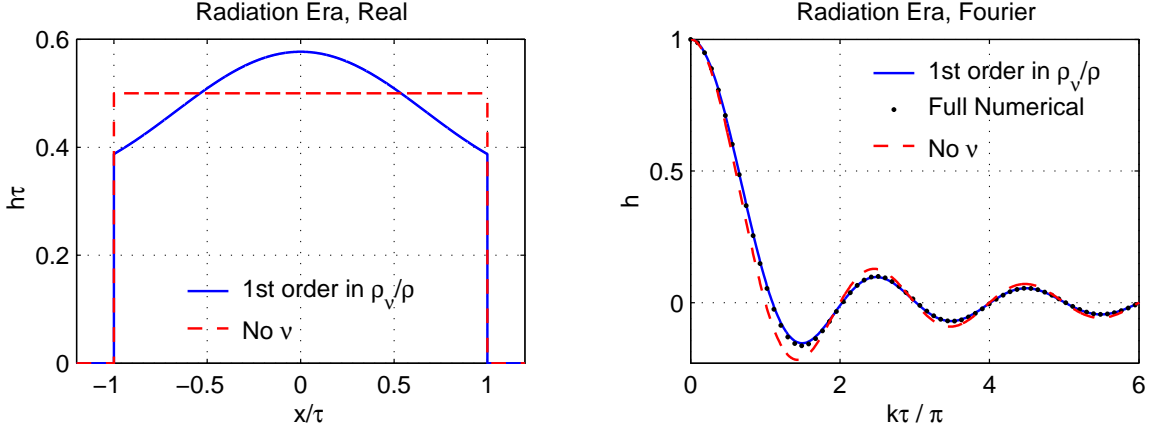}
\caption{Evolution of tensor metric perturbations
in the radiation era
as described by a real-space Green's function, left,
and Fourier modes, right. 
For both presentations, the displayed solutions are the neutrinoless,
unsourced by any anisotropic stress (dashed)
and the linear in $\rho\snu/\rho$ for three neutrino flavors (solid). 
The  Fourier modes on the right panel are additionally compared with
the full Boltzmann {\sc cmbfast} calculation (bold dots).  
}
\lb{fig_rad}
\end{figure*}

The amplitude and phase of the transfer function
oscillations in $k\t$ are fixed fully by the Green's
function discontinuity at $|\chi|=1$.
The $k\t\to \infty$ behavior of the Fourier 
modes\rf{asymp_mode_gen} maps to the following  
discontinuous terms in real space\ct{Jones_GenFunc,summary_of_gen_trf}:
\be
\bar h^{\rm (disc)}(\chi) = \fr{A_0}{2}[\cos\f_0\, \theta(1-|\chi|)
              -{\sin\f_0 \ov \pi}\,\ln|1-\chi^2|]. ~~
\lb{h_asym}
\ee
Regardless of the continuous part
of~$\bar h(\chi)$, $\bar h\equiv0$ for $|\chi|>1$,
as required by causality,
only if $\sin\f_0=0$, \ie\ if the logarithmic term in eq.\rf{h_asym}
vanishes identically. 
Appendix~B confirms that the explicit solution of 
eq.\rf{sseq} has this property for any $\Rnu$.

Since $\sin\f_0=0$, either $\f_0=0$ or $\f_0=\pi$. 
If both $A_0$ and $\f_0$ in eq.\rf{asymp_mode_gen}
change continuously\footnote{
 If we demand that the amplitude and phase change continuously in $R_\nu$,
 we should, in principle, allow the amplitude
 to become negative.
 The following calculations show that in a realistic range of~$R_\nu$
 the amplitude varies moderately and remains 
 positive.
} 
in $R_\nu$ and $\f_0\equiv0$ at $\Rnu=0$
then $\f_0=0$ for any $\Rnu$.

\subsection{First-order solution}
\lb{sec_grf_1st}

We look for an explicit Green's function
$\bar h(\chi)$ as a series in the powers of $\Rnu$.
In the zeroth order $\bar h$ is described by a top-hat function\rf{h0}.
The $O(\Rnu)$ order is addressed next;
it will match the full numerical
Boltzmann evolution to a few percent.
Since $\bar h\equiv 0$ for $|\chi|>1$,
only the interval $\chi\in[-1,1]$ is considered.

Substitution of the zeroth-order $\bar h$
into the right hand side of eq.\rf{sseq} gives
for the $O(\Rnu)$ correction 
\be
&&\lf[(\chi^2-1)\bar h^{(1)}{}'\rt]' \,= 
\lb{1st_ord_dd}\\
&&=\,\fr{3\Rnu}{2}\lf[2\chi^4-\fr{11}{3}\chi^2+1
              +(\chi^2-1)^2\chi\ln\fr{1-\chi}{1+\chi}
      \rt].
\nn
\ee
After one integration and division of both sides of the
obtained equation 
by $\chi^2-1$,
\be
\bar h^{(1)}{}'=
  \Rnu\lf[\chi\lf(\fr{\chi^2}{2\,}-1\rt)
     +\fr14(\chi^2-1)^2\ln\fr{1-\chi}{1+\chi}\rt]
\lb{1st_ord_d}
\ee
where $\chi\in(-1,1)$.

As discussed at the end of Sec.~\ref{ss_self_sim},
the division by  $\chi^2-1$ creates
delta-function spikes in $\bar h^{(1)}{}'$ 
at $\chi=\pm1$.  The delta-function prefactors 
are determined
by observing that the spikes control
the constant term in $\bar h^{(1)}=\int \bar h^{(1)}{}'d\chi$
for $\chi \in [-1,1]$.
This term
is to be fixed by $\bar h$ normalization\rf{ss_norm}, 
by which
\be
\int_{-1}^{1}\bar h^{(1)}(\chi)=0.
\lb{sum_rule1}
\ee

Integrating eq.\rf{1st_ord_d} and fixing
the constant term as described
we find that
\be
\bar h^{(1)}&=&
  \fr{\Rnu}{60}\lf\{
     6\chi^4-23\chi^2+\fr13+16\ln2\,+\rt.
\lb{h1}
\\
     &+&\!\!\lf.\lf(3\chi^4-10\chi^2+15\rt)\chi\ln\fr{1-\chi}{1+\chi}
       -8\ln(1-\chi^2)
     \rt\}
\nn
\ee
for $\chi\in[-1,1]$ and $\bar h^{(1)}=0$ otherwise.

As anticipated, $\bar h^{(1)}$ discontinuity at $|\chi|=1$
is of the form\rf{h_asym},
where $\f_0=0$ and
\be
A_0^{(1)}=\lim_{|\chi|\to 1-0} 2\bar h^{(1)} = -\,\fr59\,\Rnu.
\ee
The obtained $A_0$ determines the subhorizon radiation era solution 
for tensor modes, eq.\rf{asymp_mode_gen}, as
\be
h(k\t)=\lf(1-\fr59\,\Rnu+O(\Rnu^2)\rt)\fr{\sin k\t}{k\t}
         + O((k\t)^{-2}).\ 
\lb{h_rad_fin}
\ee
We also note that the leading behavior
of the $O((k\t)^{-2})$ correction follows from the
general result\rf{h_rad_sublead_a},
\be
h(k\t)=
        A_0 \lf\{\fr{\sin k\t}{k\t}-\Rnu\fr{\cos k\t}{k^2\t^2}
        \lf[1+o(1)\rt]\rt\},
\lb{h_rad_sublead}
\ee
valid in the radiation era at all orders of 
the $\Rnu$ expansion.

It is possible to find the entire  radiation era
evolution of a tensor Fourier mode $h(k\t)$
in $O(\Rnu)$ order. 
Introducing $\f\equiv k\t$, we calculate 
\be
h^{(1)}(\f)=\int_{-1}^1d\chi\, \bar h^{(1)}(\chi)\,e^{-i\f\chi},
\ee
with $\bar h^{(1)}(\chi)$ given by eq.\rf{h1}.
The result is 
\be
  h^{(1)}(\f)
  =\fr{\Rnu}{\f}\lf(A^{(1)}\sin\f - B^{(1)}\cos\f\rt),
\lb{h1k}
\ee
where 
\be
 \ba{l}
A^{(1)}=-\fr59-\fr1{\f^2}+\fr{24}{\f^4}+\fr{12}{\f^4}\Ci(2\f)
-4\lf(\fr1{\f^3}-\fr{3}{\f^5}\rt)\Si(2\f),\\
B^{(1)}= \fr1{\f}+\fr{12}{\f^3}+\fr{12}{\f^4}\Si(2\f)
  +4\lf(\fr1{\f^3}-\fr{3}{\f^5}\rt)\Ci(2\f).\Vsp
 \ea
\nn
\ee
Here, $\Ci x\equiv \int_0^{x}\fr{\cos y-1}{y}\,dy
 =-\ln x - \gamma + \ci x$ and 
$\Si x\equiv \int_0^{x}\fr{\sin y}{y}\,dy
     ={\pi\ov2}+\si x$.
Explicit verification with Mathematica 
confirms that in the $O(\Rnu)$ order the obtained 
 $h^{(0)}+h^{(1)}$ satisfies Weinberg's integro-differential 
equation for a tensor Fourier mode, eq.~(21) in Ref.\ct{Weinb_nu}.

Figure~\ref{fig_rad} shows the neutrinoless (dashed)
and the linear in $\Rnu$ (solid) 
gravity wave solutions in real and Fourier presentations.
Here, $\Rnu$ is $0.405$, corresponding to the
three standard neutrinos.
The  Fourier modes are additionally compared with
a full Boltzmann calculation with {\sc cmbfast}\ct{cmbfast} 
(dots).  As seen, the $O(\Rnu)$ analytical solution
provides a reasonable accuracy for all $k\t$.

For the standard $\Rnu$ value, 
the $O(\Rnu)$ tensor amplitude
suppression factor is
$A_0^{}= 1-\fr59\Rnu \simeq 0.77$.
It is in excellent agreement with the {\sc cmbfast} calculation, 
giving $A_0^{\rm (CMBFAST)}\simeq 0.76$ during radiation domination
(redshift $z=10^7$, the same result at $z=10^8$),
and with numerical solutions of Weinberg's
integro-differential equation in Fourier space,
$A_0^{\rm (Weinb)}\simeq 0.80$\ct{Weinb_nu} and 
$A_0^{\rm (PK)}\simeq 0.81$\ct{PritchKam04}.

\section{To the matter era}
\lb{sec_mat}

Most of the detectable gravity waves effects
are generated after the radiation era.
The modes 
probed by interferometers and pulsar timing
enter the horizon in the early radiation era,
and their subsequent evolution
is described excellently by 
the adiabatic approximation $h\propto \sin(k\t)/a$. 
The modes probed by CMB anisotropy and polarization 
enter when radiation and matter densities are comparable,
and the post-radiation era evolution
of these modes is more subtle.
Note that their CMB imprints are left
primarily at redshifts at which
dark energy plays little role in the 
background expansion.

\subsection{Minimal analytic solution}
\lb{ss_min}

For the studied redshifts, 
when dark energy density can be assumed small,
we describe the Hubble expansion by a radiation-matter model
with $\rho=[\rho_{\g,0}/(1-\Rnu)]/a^4+\rho_{m,0}/a^3$.
The corresponding scale factor~$a$, governed by the Friedmann 
equation, evolves as 
\be
a = \lf(2\bar\t+\bar\t^2\rt) a\eq.
\lb{a_rad_mat}
\ee
Here, $a\eq=[\rho_{\g,0}/(1-\Rnu)]/\rho_{m,0}$ 
is the value of~$a$ at 
radiation-matter equality, and $\bar\t$ is 
conformal time 
in units of characteristic conformal 
time of equality
\be
\bar\t\equiv {\t/\t_e}, \qquad
\t_e&\equiv& 2\sqrt{\frac{a_{\rm eq}}{H_0^2\Om_m}}.
\lb{t_e_def}
\ee
Note that by eq.\rf{a_rad_mat} 
$\t(a_{\rm eq}) \equiv \t_{\rm eq} = (\sqrt2 - 1)\t_e$.

First, we address the modes
that enter the horizon 
long before equality ($k\gg\t\eq^{-1}$)
although after neutrino decoupling.
After a mode has become subhorizon ($k\gg\H$),
we ignore the anisotropic stress.
Then the mode evolves according to 
a homogeneous wave equation
\be
\ddot h + 2\H\dot h + k^2 h = 0. 
\ee
The Hubble redshift  $2\H\dot h$ 
of the graviton energy is accounted for
by substitution 
$
h={\const\ov a}\,\v,
$
where now\ct{Grishch74}
\be
\ddot \v + \lf(k^2-\fr{\ddot a}{a}\rt) \v = 0. 
\lb{ddot_f_a}
\ee
In the small-scale limit $k/\H\to \infty$, 
the term $\ddot a/a\sim \H^2$ can be dropped.  In this (adiabatic) 
approximation $\v$~oscillates harmonically in
$\f=k\t$ with constant amplitude.
Correspondingly, the amplitude  of $h$ oscillations
decays as $1/a$.

We normalize the amplitude and fix the phase shift of 
the considered short-scale modes
by subhorizon radiation era solution\rf{h_rad_fin}.  
As a result,
\be
h(\t,k\to\infty)=A(\t)\,h^{\rm(rad,0)}(\f),
\lb{large_k}
\ee
where 
\be
h^{\rm(rad,0)}=j_0(\f)=\fr{\sin \f}{\f},
\lb{trf_rad}
\ee
is the neutrinoless radiation era solution,
and $A(\t)\propto \f/a$ equals
\be
A(\t)=\fr{1-\fr59\,\Rnu}{1+\fr12\bar\t}.        
\lb{A_t}
\ee
The last formula neglects the $O(\Rnu^2)$ 
correction in eq.\rf{h_rad_fin} and 
uses $a(\t)$ form\rf{a_rad_mat}.
We note that the Fourier
short-scale evolution\rf{large_k} 
maps onto the following Green's function 
discontinuity, \cf eqs.\rfs{asymp_mode_gen}{h_asym}:
\be
h^{\rm (disc)}(\t,x)={A(\t)\ov 2\t}\,\theta(\t-|x|)
                 =A(\t)\,h^{\rm(rad,0)}(\t,x).\ 
\lb{disc}
\ee

Next, we consider the opposite, large-scale limit ($k\ll\t\eq^{-1}$),
in which mode evolution starts only in the matter era.
Then anisotropic stress of relativistic species is negligible 
and, similarly to the radiation era,
the tensor dynamics is scale-free.
Solving eq.\rf{ddoth}, where now $\Sigma=0$ and $\H=2/\tau$, 
with a normalized Green's function of the form\rf{h_selfsim},
we obtain
\be
h^{\rm (mat,0)}={3\ov 4\t}\lf(1-{x^2\ov\t^2}\rt)\theta(\t-|x|).
\lb{grf_mat}
\ee
Fourier transforming, we arrive at a well-known tensor
transfer function in a matter dominated universe
\be
h^{\rm(mat,0)}=3j_1(\f)/\f=3\lf(\fr{\sin \f}{\f^3}-\fr{\cos \f}{\f^2}\rt).
\lb{h_mat0}
\ee

Finally, we combine the Green's function discontinuity\rf{disc}, 
describing the short modes, with the matter era 
evolution\rf{grf_mat} on the
larger scales
by a ``minimal'' analytic formula.
Namely, we look for an expression 
of the form $(a+bx^2)\,\theta(\t-|x|)$ 
that is normalized ($\int h\,dx=1$) and
jumps to  $A(\t)/(2\t)$ at the horizon $|x|=\t$. 
Such an expression is unique and is given by 
\be
h\sim A(\t)\,h^{\rm(rad,0)}+[1-A(\t)]\,h^{\rm(mat,0)}.
\lb{trf_gen}
\ee

If we view formula\rf{trf_gen}  
as a transfer function $h(\t,k)$
in Fourier space, 
with $h^{\rm(rad,0)}$\rf{trf_rad} and $h^{\rm(mat,0)}$\rf{h_mat0}, 
we observe that $h$ approaches unity for either $\t\to 0$ or $k\to 0$.
This formula
 also reproduces the correct, in $O(\Rnu)$,
asymptotic behavior for $k\to \infty$
at any fixed $\t$.

Figure~\ref{fig_glob} compares the approximation\rf{trf_gen}
(solid curve)
with numerical {\sc cmbfast}\ct{cmbfast} calculation
(bold dots) at CMB last scattering, $z=1090$. 
The heights of all the oscillation peaks match
within $11\%$ and the phase within $3\%$ of $\pi$.
Neither the radiation-era transfer function $A(\t)\,h^{\rm(rad,0)}$,
corrected for the neutrino and adiabatic damping (dashed) 
nor the popular matter-era transfer function
$h^{\rm(mat,0)}$ (dash-dotted)  
alone can describe the tensor
evolution to CMB last scattering satisfactorily 
for all~$k$. 

The good agreement between minimal analytic formula\rf{trf_gen}
and numerical calculations is preserved at later times,
which are relevant for tensor  signatures
in ISW distortion of CMB temperature
and large-scale polarization from reionization 
(reionization bump). 
A comparison for a redshift $z=109$ is given
in Fig.~\ref{fig_glob2}.

\begin{figure}[tb]
\includegraphics[width=7cm]{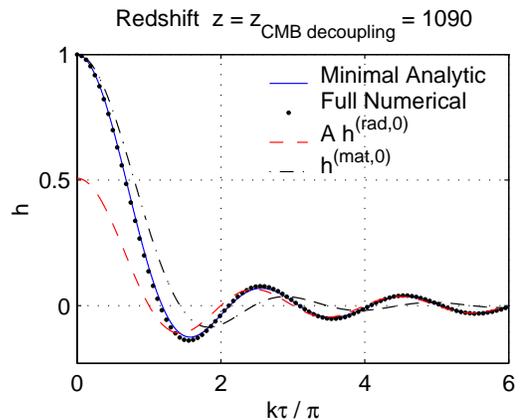}
\caption{Tensor modes at the redshift of CMB last scattering, $z=1090$. 
The plots are generated with: 
``minimal'' analytical formula\rf{trf_gen} (solid),
numerical {\sc cmbfast} calculation (bold dots),
the neutrino- and adiabatically-damped radiation era 
solution (dashed), and the matter era solution (dash-dotted).
}  
\lb{fig_glob}
\end{figure}
\begin{figure}[tb]
\includegraphics[width=7cm]{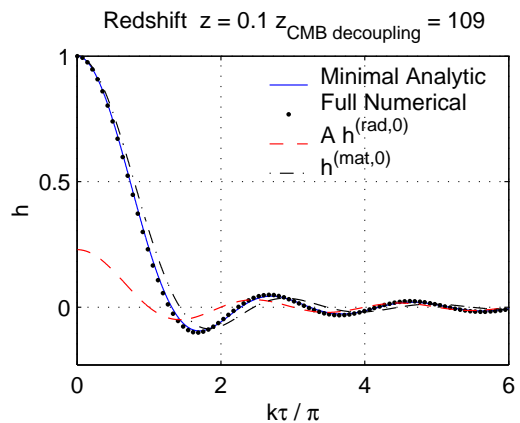}
\caption{The same calculations as in Fig.~\ref{fig_glob} 
but for a ten times smaller redshift.  ``Minimal''
analytical expression\rf{trf_gen} (solid) 
continues to provide good accuracy. 
}  
\lb{fig_glob2}
\end{figure}

We highlight that approximation\rf{trf_gen} preserves 
the temporal phase shift of mode oscillations 
after radiation-matter equality.
On subhorizon scales,
this formula reduces for $\t\gg \t\eq$ to
\be
h^{\rm(sub,\, mat)}\sim {2k\t_e(1-\fr59\,\Rnu)\sin \f -3\cos\f \ov \f^2}.
\lb{hmin_mat_subhor}
\ee
As the original formula\rf{trf_gen}, 
asymptotic expression\rf{hmin_mat_subhor}
has the correct leading-order behavior
in both $k\t_e\to 0$ and $k\t_e\to \infty$ limits.
However, the actual subleading corrections to these limits
differ from the terms of eq.\rf{hmin_mat_subhor}
in magnitude and, when $k\t_e\to \infty$, even in $k$~scaling.
The evolution of a tensor mode with a finite~$k$ is studied
next quantitatively.

\subsection{Entering before equality ($k\t\eq>1$)}
\lb{ss_rad_ent}

First, to determine the role of the radiation-matter transition
in mode evolution, we ignore neutrino anisotropic stress.
The derived formula will ultimately be corrected for it.
Without anisotropic stress, an exact formal
expression for tensor evolution in the radiation-matter background
can be given in terms of radial prolate spheroidal 
functions\ct{Sahni88_h_analyt,KorAll94_h_analyt}.
Yet, a cumbersome form of this expression (an infinite series
of Hankel functions with coefficients involving more
infinite series), calls for more
insightful and practical description.

We continue to parameterize tensor mode evolution by the dimensionless
oscillation phase $\f\equiv k\t$ and use the 
perturbation variable $\v\propto ah$.
We normalize $v$ by $\v(\f)=\f+o(\f)$ as $\f\to 0$.
Then the initial condition $h(\t\to0)=1$
and explicit $a(\t)$ solution\rf{a_rad_mat}
fix $ah/v$ to
\be
h=\fr{2a\eq}{a\f_e}\,\v= \fr{\v}{\lf(1+\fr12\bar\t\rt)\f}\,.
\lb{h-v}
\ee
In the first of these equalities and later
\be
\f_e\equiv k\t_e.
\ee

Source-free dynamical equation\rf{ddot_f_a} for $\v(\f)$
in the radiation-matter background reads 
\be
\v'' + \lf(1-\fr1{\f\f_e+\fr12\f^2}\rt) \v = 0.
\lb{fprime}
\ee
Here and below primes denote derivatives with
respect to~$\f$.
In a purely radiation background, 
in which $\ddot a=0$ and $\v'' +\,\v = 0$,
the solution is $\v^{\rm (rad,0)}=\sin\f$.
To account for massive matter  
{\it deep in the radiation era\/} ($\f \ll\f_e$)
we treat $(\ddot a/a)\v$ in eq.\rf{ddot_f_a} 
perturbatively, approximating it by
$(\ddot a/a)\v^{\rm (rad,0)}$. 
Then, neglecting $\f^2$ against 
$\f\f_e$, we have
\be
\v'' + \v \approx \fr{\sin\f}{\f\f_e}.
\lb{fprime_rad_pert}
\ee

A substitution $\v = \A(\f)\sin\f$
leads to a first-order differential equation for $\A'(\f)$.
That equation has a unique regular at $\f=0$ solution
\be
\A' = \fr{1}{2\f_e\,{\sin^2}\f}\lf[\ln(2\f)+\gamma - \ci(2\f)\rt],
\lb{dA_sol}
\ee
where $\gamma=0.577216...$ is the Euler constant 
and $\ci x\equiv-\int_x^{\infty}\cos y\,dy/y$
is the cosine integral.
Integrating eq.\rf{dA_sol}, we obtain:
\be
\v^{\rm (rad)}&=&A(\f)\sin\f-B(\f)\cos\f,
\lb{osc_gen_real}
\ee
where
\be
A&=&1+\fr1{2\f_e}\lf[\fr{\pi}{2}+\si(2\f)\rt],
\lb{A_rad_ent}\\
B&=&\fr1{2\f_e}\lf[\ln(2\f)+\gamma - \ci(2\f)\rt],
\lb{B_rad_ent}
\ee
and $\si x\equiv-\int_x^{\infty}\sin y\,dy/y$
is the sine integral.

{\it Well under the horizon\/} $(\f\gg1)$, the sine and 
cosine integrals in expressions\rfs{A_rad_ent}{B_rad_ent} tend to zero.
Then
\be
\v^{\scriptstyle\rm rad,\choose \scriptstyle\rm sub}\! 
      \approx \lf(1+\fr{\pi}{4\f_e}\rt)\sin\f
       -\fr1{2\f_e}\lf[\ln(2\f)+\gamma\rt]\cos\f.~
\lb{rad_sol_subhor}
\ee
We see that at a large but finite $\f_e\equiv k\t_e$
the oscillation phase is shifted
from the $\sin\f$ oscillations 
of the adiabatic limit.
The shift is  caused by nonrelativistic matter 
contributing to the 
background expansion.  As the 
matter role increases in time, the shift 
grows logarithmically even after the tensor mode has become subhorizon.

We can easily extend the mode subhorizon evolution
{\it beyond the radiation era\/} using the WKB 
approach\footnote{
  Since the adiabatic decay $h\sim 1/a$
  on subhorizon scales is already accounted for by the  $\v$
  prefactor, \eg\ eq.\rf{h-v}, the presented first-order 
  WKB calculation of $\v$ is equivalent to the 
  {\it second\/}-order WKB analysis of the tensor metric perturbation $h$.}.  
To that end, we combine the prefactors $A$ and $B$ of eq.\rf{osc_gen_real}
into a varying complex amplitude $C\equiv A-iB$:
\be
\v = \Ip \lf(C\,e^{i\f}\rt).
\lb{WKB_ansatz}
\ee
$C(\f)$ satisfies a 
second-order differential equation, following from eq.\rf{fprime}.
After the horizon entry $C(\f)$ changes
slowly, hence, we neglect $C''(\f)$ in that equation.
The remaining terms are 
\be
C'=\fr{-i}{2\f\f_e+\f^2}\,C.
\lb{eq_WKB}
\ee
Integrating\rf{eq_WKB} and 
setting the integration constant to match 
the mode\rf{WKB_ansatz} 
to radiation era solution\rf{rad_sol_subhor}, 
we find
\be
\v^{\rm (sub)}=\lf(1+\fr{\pi}{4\f_e}\rt) 
    \sin(\f-\D\f),
\lb{WKB_sol}
\ee
with
\be
\D\f^{\rm (sub)}=\fr1{2\f_e}\lf[-\ln\lf(\fr1{4\f_e}+\fr1{2\f}\rt)
          +\gamma\rt] 
\lb{Df_sub}
\ee

Finally, we incorporate the damping by neutrinos
during the horizon entry
in the radiation era.
We treat the correction caused by neutrino anysotropic stress 
as perturbation which is independent of the perturbation
induced by the matter term
in eq.\rf{fprime_rad_pert}.
In this approximation, neglecting interference
of the two corrections, neutrinos additionally
suppress the oscillation amplitude by
the factor in eq.\rf{h_rad_fin} 
but they do not affect the oscillation phase.
Adjusting the $\v^{\rm (sub)}$~amplitude  
in eq.\rf{WKB_sol} and substituting the result 
into eq.\rf{h-v}, we thus have:
\be
h=\fr{A(\t,k)}{\f}\,\sin[\f-\D\f(\t,k)],
\lb{h_sub_gen}
\ee
where on subhorizon scales
\be
A^{\rm (sub)}\approx \fr{1-\fr59\,\Rnu+\fr{\pi}{4\f_e}}{1+\fr12\bar\t}   
\lb{A_sub_radent}
\ee
and $\D\f$ is given by eq.\rf{Df_sub}.

{\it Deep in the matter era\/}
($\f\gg\f_e$), the obtained~$A^{\rm (sub)}$ decays inversely with time as
\be
A^{\rm (sub,\,mat)}\approx 
  \fr{2\f_e(1-\fr59\,\Rnu)+ {\pi\ov2}}{\f},
\lb{A_sub_mat}
\ee
while the $\D\f^{\rm (sub)}$ growth saturates at 
\be
\D\f^{\rm (sub,\,mat)}\approx  \fr{\ln(4\f_e)+\gamma}{2\f_e}.
\lb{Df_sub_mat}
\ee
Since $\f_e= k\t_e$, the phase shift of tensor oscillations 
at fixed time and $k\to\infty$ approaches
its asymptotic zero value as $\ln k/k$, more slowly than
could be expected from naive Taylor expansion. 
We recall the origin of this scaling in
the impact of matter density, which fraction increases in time and 
induces logarithmic growth of the tensor phase shift
before radiation-matter equality.

\begin{figure}[tb]
\includegraphics[width=7cm]{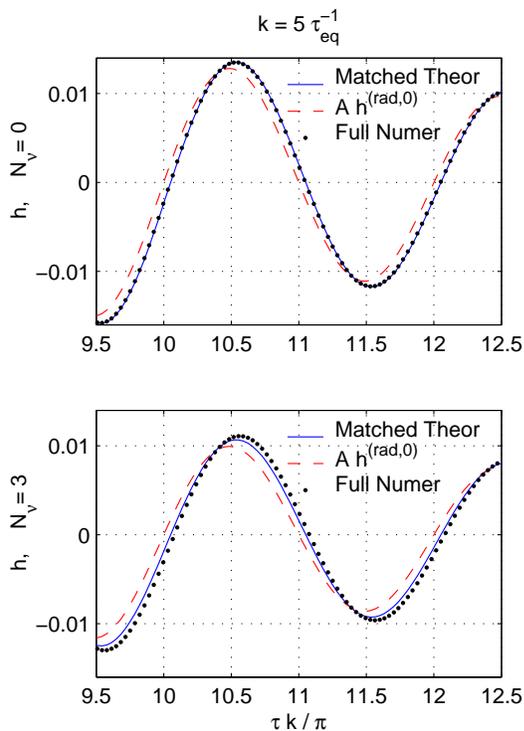}
\caption{
The sixth oscillation of a tensor mode with $k=5\t\eq^{-1}$
as:
next-to-adiabatic analytical 
solution~(\ref{h_sub_gen})-(\ref{A_sub_radent})-(\ref{Df_sub})
(solid), {\sc cmbfast} output (dots), and
the neutrino- and adiabatically-damped radiation era 
solution (dashed). The top figure is for the $\rho\snu=0$ model. 
Both the matter-induced 
amplitude enhancement and phase shift
are seen well reproduced analytically.  
The bottom figure describes the standard 3-neutrino scenario.
The appearing small discrepancy is caused by the interference
of the neutrino and matter corrections, as discussed in the main text.
}  
\lb{fig_rad_ent}
\end{figure}

The accuracy of our 
result~(\ref{h_sub_gen})-(\ref{A_sub_radent})-(\ref{Df_sub})
against numerical calculations is seen
from Figs.~\ref{fig_rad_ent}, which capture the sixth
oscillation cycle of a tensor mode with $k\t\eq=5$.
The plotted are: the neutrino- and adiabatically-damped radiation era 
solution (dashed), our next order analytical prediction (solid),
and the full {\sc cmbfast} calculation (dots). 
The top Figure describes the $\rho\snu=0$ case
and shows the excellent accuracy of the next-to-adiabatic 
analytical predictions for both the amplitude enhancement
and phase shift induced by matter.  
The bottom Figure refers to the standard 3-neutrino 
scenario. 
Then, in addition to the neutrino and matter
effects, their unaccounted interference 
slightly increases the amplitude and the phase shift.
The higher amplitude is expected from dilution of neutrinos 
by matter.  
The larger phase shift may be understood
as follows: 
in the radiation era according to  eq.\rf{h_rad_sublead}
neutrino anisotropic stress induces a decaying 
shift of oscillations toward higher~$\f$. When the neutrino 
source is cut off by matter domination at a finite 
$\f\sim \f_e$, the oscillations remain 
shifted by $\D\snu\f=O(\Rnu/\f_e)$.

\subsection{Entering after equality ($k\t\eq<1$)}
\lb{ss_mat_ent}

Now we address the modes that start to evolve after
radiation becomes subdominant to nonrelativistic matter.
We determine the tensor evolution by retaining only the leading, 
linear in $\f_e=k\t_e$, corrections due to the residual effects 
of radiation. 
We could treat the 
radiation-induced corrections perturbatively,
similarly to treating the subdominant 
matter effects in the previous subsection.
However, the derivation becomes more elegant and 
the result is easier to interpret if, instead, 
we account for the radiation energy
by a simple change of variables, as described next.

We start by rewriting the scale factor evolution\rf{a_rad_mat}
as
\be
a = a\eq\lf[(\t+\t_e)^2-\t_e^2\rt]/\t_e^2 .
\ee
In terms of the evolution variable $\f=k\t$,
\be
a = {a\eq \ov \f_e^2}\lf(\~\f^2-\f_e^2\rt),
\lb{a_shifted}
\ee
where
\be
\~\f\equiv k(\t+\t_e)= \f+\f_e.
\ee
Next, in eq.\rf{a_shifted} we drop the 
last $\f_e^2$ term and in the gravitational wave eq.\rf{ddoth_i}
ignore the anisotropic stress of radiation, both being $O(\f_e^2)$ corrections.
Now, with $a(\t)\propto \~\f^2$, 
we observe that all the remaining contribution of radiation  
to eq.\rf{ddoth_i} 
can be accounted for by considering 
a matter-only scenario but
replacing $\f$ by the shifted evolution variable~$\~\f$, 
whose range is $(\f_e,+\infty)$.
In other words, a solution of full
eq.\rf{ddoth_i} can be written as
\be
h(\t,k)
   = \bar h^{(\rm mat,0)}(\~\f)+O(\f_e^2),
\ee
where $\bar h^{(\rm mat,0)}$ is a matter-only 
solution with so far unspecified initial conditions.

Although the omission of the last $\f_e^2$ in 
eq.\rf{a_shifted} has a big impact on $\H$
when $\t\lesssim \t_e$, the considered large-wavelength 
perturbation is then frozen, whether we apply
the true or modified Hubble expansion.  
Therefore, as can be easily verified, 
the $\f_e^2$ omission leads to a minor
$O(\f_e^2)$ change in the mode evolution.

In view of the last remark, we continue to impose
the initial conditions $h=1$ and $h'=0$
at $\f=0$, \ie, at $\~\f=\f_e$.  These conditions 
fix the constants $C_1$ and $C_2$ in the general solution
\be
h(\t,k)\approx C_1 h^{(\rm mat,0)}(\~\f) + C_2\,n(\~\f),
\lb{h_mat_superpos}
\ee
where $h^{(\rm mat,0)}$ is the regular matter era solution\rf{h_mat0}
and $n$ is a complimentary linearly independent solution.
To be specific, we take
\be
n(\f)= 3\lf(\fr{\cos \f}{\f^3}+\fr{\sin \f}{\f^2}\rt).
\lb{mat_sol_compl}
\ee
Then $h^{(\rm mat,0)}$ and $n$ are comparable on subhorizon scales.  
Given that $h^{(\rm mat,0)}(\f_e)= 1+O(\f_e^2)$ and
$n(\f_e)=O(\f_e^{-3})$, the initial conditions
fix the weights in eq.\rf{h_mat_superpos} as
\be
C_1=1+O(\f_e^2),\qquad C_2=O(\f_e^5).
\ee
We conclude that up to $O(\f_e^2)$ corrections
\be
h(\t,k)\approx h^{(\rm mat,0)}(\~\f) 
      = 3\lf(\fr{\sin \~\f}{\~\f^3}-\fr{\cos \~\f}{\~\f^2}\rt).
\lb{h_mat1}
\ee
This result can also be derived by a regular
perturbative in $\f_e$ analysis.

We thus found that in $O(\f_e)$ order
the residual radiation density pushes
the tensor modes ahead of the zeroth-order
solution $h^{(\rm mat,0)}(k\t)$ by a constant conformal time 
increment~$\t_e\simeq 260$\,Mpc. 
 Fig.~\ref{fig_mat_ent} shows good
agreement between eq.\rf{h_mat1} (solid)
and {\sc cmbfast} integration (bold dots) for 
a mode with $k=0.2\,\t\eq^{-1}$.
\begin{figure}[tb]
\includegraphics[width=6.7cm]{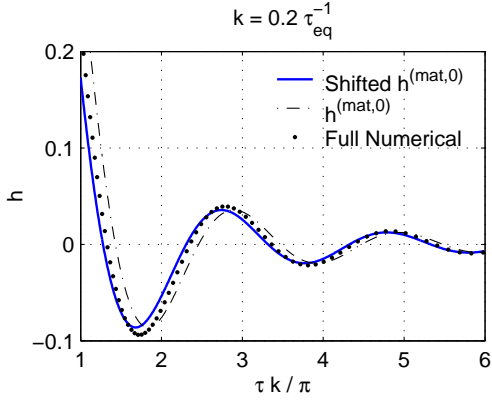}
\caption{
Evolution of a mode with $k=0.2\,\t\eq^{-1}$ as:
the shifted $\t\to\t+\t_e$  matter era solution (solid),
the regular matter era solution (dash-dotted),
and {\sc cmbfast} integration (dots).
}  
\lb{fig_mat_ent}
\end{figure}

In the subhorizon limit, eq.\rf{h_mat1} reduces to
\be
h^{\rm(mat,\, sub)} \approx -3\cos(\f+\f_e)/\f^2.
\ee
The amplitude suppression factor $A$ and the phase shift $\D\f$
of our earlier parameterization $h=A(\t)\sin[\f-\D\f(\t)]/\f$
are now equal 
\be
A^{\rm (mat,\, sub)}\approx \fr3{\f},\qquad
\D\f^{\rm (mat,\, sub)}\approx \fr{\pi}{2}-\f_e.
\ee
These values 
for the large wavelengths
can be compared with results\rf{A_sub_mat} and\rf{Df_sub_mat},
derived for the scales that enter the horizon before equality.
If desired, these two sets of asymptotic formulas can be
joined by various fitting ansatzs for
a more accurate interpolation than provided by 
our minimal formula\rf{hmin_mat_subhor}.

\appendix

\section{Free streaming in general relativity}
\lb{gen_free_stream}

\subsection{Full theory}

Classical particles can be described by a phase space distribution~$f$
over spacetime coordinates
$(\t,x^i)$ and particle {\it canonical\/} momenta~$P_i$.
We consider $P_i$, as opposed to more conventional\ct{BondSz83,MaBert95,Dod_book} 
comoving proper momenta, 
as the primary phase space coordinates.
At a minor expense of 
two extra terms in the linearized equations
for the energy-momentum tensor,
eqs.\rf{T00pert_a} and\rf{Sigma_gen_a},
our choice leads, first, to 
a simpler general-relativistic equation 
of $f$ evolution. Second,
to time independence of $f$
for free particles on superhorizon scales\ct{B04}.

The phase space distribution of free streaming particles
evolves according to the collisionless Boltzmann equation
\be
\dot f + \fr{dx^i}{d\t}\,\fr{\pd f}{\pd x^i}
       + \fr{dP_i}{d\t}\,\fr{\pd f}{\pd P_i} = 0.
\lb{f_dot_gen_a}
\ee
The canonical momentum of a particle that has mass $m$ 
and moves along a worldline $x^\mu(\t)=(\t,x^i(\t))$ is 
\be
P_i = m\,{g_{i\mu}dx^{\mu}}/(-ds^2)^{1/2},
\ee
where $ds^2=g_{\mu\nu}dx^{\mu}dx^{\nu}$.
Natural definitions $P_0 \equiv m\,{g_{0\mu}dx^{\mu}}/(-ds^2)^{1/2}$, 
$P^\mu \equiv g^{\mu\nu}P_\nu$ and the calculation of $dP_i/d\t$
from the geodesic equation give
\be
\fr{dx^i}{d\t} = \fr{P^i}{P^0},\qquad
\fr{dP_i}{d\t} = \fr{g_{\mu\nu,i}P^{\mu}P^{\nu}}{2P^0}.
\lb{Fi_def_a}
\ee
The value of $P_0$ 
at every phase space point $(x^i,P_i)$ 
follows from the identity
\be
-g^{\mu\nu}P_\mu P_\nu= m^2.
\lb{on_shell_a}
\ee

We specify the direction of particle
propagation by the spatial components of
\be
n_\mu\equiv \fr{P_\mu}{P},\quad
\mbox{where}\quad 
P^2\equiv \sum_{i=1}^3 P_i^2.
\lb{Pn_def_a}
\ee
This metric-independent, therefore non-covariant,
definition of $n_\mu$ 
separates the infinitesimal transport of 
particles from the temporal change of metric.
Note that by eq.\rf{Pn_def_a}
\be
\sum_{i=1}^3 n_i^2=1.
\lb{sum_ni_a}
\ee
We also consider $n^\mu\equiv g^{\mu\nu}n_\nu = P^\mu/P$.

The energy-momentum tensor of the described particles equals
\be
T^{\mu }_{\nu }
   =\int\fr{d^3P_i}{\rtg}\fr{P^\mu P_\nu}{P^0}\,f.
\lb{Tmunu_gen_a}
\ee
We substitute $P_\mu=n_\mu P$ and $P^\mu=n^\mu P$,
and define particle {\it intensity\/}~$I$
in a direction~$n_i$ as
\be
I(x^\mu,n_i)\equiv \int_0^{\infty} P^3dP\,f(x^\mu,n_iP).
\lb{I_def_a}
\ee
Then
\be
T^{\mu }_{\nu}
   =\int\fr{\dOmn}{\rtg}\fr{n^\mu n_\nu}{n^0}\,I.
\lb{Tmunu_int_a}
\ee
The dynamics of the intensity of {\it ultra-relativistic\/} 
decoupled particles 
is governed by a simple closed equation, derived next.

Only two of the three components of $n_i$,
constrained by condition\rf{sum_ni_a}, 
are independent. Let $\mu_\al = (\mu_1,\mu_2)$
be any two independent variables
parameterizing~$n_i$.
Then partial differentiation with respect to 
$n_i=P_i/P$ 
is naturally defined as a differential operator
\be
\fr{\pd }{\pd n_i}\equiv \sum_{\al=1,2}\lf(\fr{P\pd\mu_\al}{\pd P_i}\rt) 
    \fr{\pd}{\pd\mu_\al}.
\ee

Integrating the Boltzmann equation\rfs{f_dot_gen_a}{Fi_def_a}
over $P^3dP$, applying the above definition of $\pd/\pd n_i$,
and remembering the condition of ultra-relativity 
$g^{\mu\nu}n_\mu n_\nu=0$, we obtain
\be
n^\mu\fr{\pd I}{\pd x^\mu}= \fr12\,n^\mu n^\nu g_{\mu\nu,i}
     \lf(4n_iI-\fr{\pd I}{\pd n_i}\rt).
\lb{dotI_gen_a}
\ee
Equation\rf{dotI_gen_a} is the main result of this Appendix.

As this dynamical equation shows,  for superhorizon perturbations
the considered intensity of decoupled particles
is time independent. 
Indeed, when the spatial gradients $\pd I/\pd x^i$ 
and $g_{\mu\nu,i}$ are negligible, 
eq.\rf{dotI_gen_a} becomes $\dot I=0$.
In the opposite extreme, for the locally Minkowski metric, 
$I$~reduces to the conventional particle intensity $dE/(dVd^2\^{\bm n})$.

\subsection{Linear theory}

For a homogeneous and isotropic distribution 
of particles with energy density  $-T^0_0 \equiv \rho(\t)$ 
we see from eq.\rf{Tmunu_int_a} that $\bar I=\rho a^4/(4\pi)$.
We describe linear inhomogeneities over this background
with relative intensity perturbation~$\di(x^\mu,n_i)$,
\be
I=\fr{\rho a^4}{4\pi}\lf(1+\di\rt).
\lb{I_lin_a}
\ee

When the particles are ultra-relativistic and 
their number is conserved ($\rho a^4$ is constant),
linearization of eq.\rf{dotI_gen_a}
gives:
\be
\dot \di + n_i\di_{,i}  
  = 2n_i\^n^\mu \^n^\nu h_{\mu\nu,i},
\lb{dotD_a}
\ee
where $\^n^\mu \equiv (1,n_i)$.
The linearly perturbed~$T^\mu_\nu$ components\rf{Tmunu_int_a} 
are straightforwardly calculated
from $n_\mu n_\nu g^{\mu\nu}=0$
and $\langle n_i \rangle_{\n}=0$,
$\langle n_in_j \rangle_{\n}=\d_{ij}/3$,
\dots, where $\la\ra_{\n}$ stands 
for $\int {\dOmn}/{4\pi}$:
\be
 -T^0_0 &=& \rho\lf(\vphantom{\dot I} 1+\la\di\ra_{\n} - 2h\rt),
\lb{T00pert_a}\\
 T^0_i &=& \la n_i\di\ra_{\n} ,\vsp \\
 T^i_j &=& \fr{-T^0_0}3\d^i_j + \Sigma^i_j.
\ee
Here, $h\equiv \fr13\sum_i h_{ii}$,
\be
\Sigma^i_j = \rho\lf[\la(n_in_j-\fr13\,\d_{ij})\di\ra_{\n} 
                     - \fr4{15}\,\hat h_{ij}\rt]
\lb{Sigma_gen_a}
\ee
is anisotropic stress,
and in the last formula $\hat h_{ij} \equiv h_{ij}-\d_{ij}h$ 
is the traceless part of $h_{ij}$.

If the energy distribution of particles that stream
in a direction $n_i$ is thermal and the metric 
is locally Minkowski then $\di$ is proportional
to the particle temperature perturbation:
$\di=4\D T(n_i)/T$.
Under a change of coordinates 
$x^\mu\to \~x^\mu=x^\mu+\al^\mu$
(gauge transformation),
$\di$ transforms as
\be
\~\di(x^\mu,n_i) - \di(x^\mu,n_i) = - 4n_in_\mu \al^\mu_{,i}. 
\lb{di_gtr_a}
\ee

Similarly to the multipoles of $\D T(n_i)/T$,
the multipoles of intensity perturbation $\di(n_i)$ can be used to decompose the transport
eq.\rf{dotD_a} into a computer-friendly Boltzmann hierarchy of equations
for photons (with scattering terms and polarization added), neutrinos, and other 
cosmological species.
These equations for scalar perturbations 
in the Newtonian gauge are given in Sec.~III and Appendix~A of\ct{B04}
(\cite{B04} uses $D\equiv\fr34\di$).
The structure of these equations is considerably simpler 
than that of the traditional description with
$\D T(n_i)/T$ multipoles: The dynamical matter perturbations are 
no longer driven by time derivatives of gravitational potentials,
which are constrained non-dynamically by the same matter perturbations.
Moreover, unlike the situation with the traditional approach, 
the change of the perturbation value during the horizon entry is 
controlled only by physical interactions and is not
caused by gauge artifacts\ct{B04}.

\subsection{Initial conditions}

We specify the initial conditions by following Ref.\ct{Weinb_nu}.
We suppose that the considered species (later, neutrinos) 
decouple at time $\t_{\rm dec}$ when the Hubble scale
is much smaller than 
the spatial scale of the probed perturbations.
Then shortly after $\t_{\rm dec}$ 
there exists a hypersurface with the following property.
Every observer on this hypersurface who moves normally to it 
would detect the same isotropic distribution of the
proper neutrino momentum~$\bm p$.
This is a hypersurface of constant temperature
when the particle distribution is thermal.
Although the decoupled relic neutrinos are not exactly 
thermal,  the described conditions should still be met on 
a hypersurface~$(u)$ of uniform neutrino energy density.
We choose $u$ for an initial constant time hypersurface.

An observer who moves normally to a constant time hypersurface 
measures proper momentum $\bm p$ which
is related to the canonical momentum~$P_i$ as
\be
p^2=\gsp^{ij}P_iP_j
\ee
($\gsp^{ij}$ is the inverse of the 3-tensor $g_{ij}$).
In terms of the variables\rf{Pn_def_a}, 
$p^2=\gsp^{ij}n_in_jP^2$.
Therefore, 
if the hypersurface of the uniform distribution
of the proper momentum
is taken for a constant time hypersurface
then, according to eq.\rf{I_def_a},
the neutrino intensity varies as
\be
I^{(u)}(\x,n_i)=\fr{\const}{\lf[n_in_j\gsp^{ij}_{(u)}(\x)\rt]^2}.
\lb{ic_gen_a}
\ee

In linear theory and the metric\rf{metric}, 
$\gsp^{ij}=a^{-2}(\d_{ij}-h_{ij})$.
Then comparing eq.\rf{ic_gen_a} and\rf{I_lin_a} 
and remembering\rf{sum_ni_a} we conclude
that on superhorizon scales the intensity 
perturbation equals
\be
\di^{\!(u)}=2n_in_jh_{ij}^{(u)}.
\lb{ic_lin_a}
\ee

\subsection{Linear solution}

Sourced transport equation\rf{dotD_a} 
with initial condition\rf{ic_lin_a} 
at $\t=\t\i$ is solved by
\be
\fr12\,\di(\t,\x,\n)&=&n_in_jh_{ij}^{(u)}(\t\i,\x\i) \,+
\lb{D_los1_a}
\\
   &&+\int_{\t\i}^{\t}\!d\t' n_i\^n^\mu \^n^\nu h_{\mu\nu,i}(\t',\x'),
\nn
\ee
where $\x\i\equiv \x+\n(\t\i-\t)$ and $\x'\equiv \x+\n(\t'-\t)$.
At the lower limit of the integral, 
$h_{\mu\nu}$~should, strictly speaking, 
be evaluated on the uniform density hypersurface
on which the initial conditions were set.
However,
for gauge-independent tensor perturbations,
 a hypersurface choice is irrelevant.
Moreover, even for scalar perturbations
the entire integral can be evaluated in
any of such traditional gauges as the synchronous, Newtonian,
comoving, or spatially flat gauges. Indeed, under a gauge
transformation $\~x^\mu=x^\mu+\al^\mu$, the integral transforms
as $-2n_in_\mu \al^\mu_{,i}(\t,\x)+2n_in_\mu \al^\mu_{,i}(\t\i,\x\i)$.
The first of these two terms matches the transformation
of the left hand side of eq.\rf{D_los1_a}. 
And the second one is negligible for 
a transformation of superhorizon perturbation
from the uniform density gauge to any of the mentioned 
gauges\ct{verify_suph_gauge_indep}.

In the integral of eq.\rf{D_los1_a}, the gradient of $h_{\mu\nu}$  
can be traded for time derivatives:
\be
n_i h_{\mu\nu,i}(\t',\x')
= \lf(\fr{d}{d\t'}-\fr{\pd}{\pd\t'}\rt)h_{\mu\nu}(\t',\x').
\nn
\ee
The trivial integration of the full time derivative gives
\be
\fr12\,\di&=&  n_in_jh_{ij}^{(u)}(\t\i,\x\i)+ 
       \lf.\^n^\mu \^n^\nu h_{\mu\nu}\rt|_{\t\i,\x\i}^{\t,\x}-
\nn\\
  &&-\int_{\t\i}^{\t}\! d\t'\, \^n^\mu \^n^\nu \dot h_{\mu\nu}(\t',\x').
\lb{D_los_a}
\ee
This solution is applicable to any (scalar, vector, or tensor) 
type of linear perturbation of ultrarelativistic species which decouple 
on superhorizon scales. 

\section{Formal proof of $\dst\lim_{k\to\infty}\f_0=0$}
\lb{appx_proof}

Sec.~\ref{sec_phase} of the main text argues
that subhorizon tensor oscillations
in the radiation era $h^{\rm(rad)}(\f)\to A_0\sin(\f+\f_0)/\f$,
$\f\equiv k\t$ are consistent with causality if only $\f_0=0$.
The unshifted oscillations  match to a finite, step
discontinuity of the real space tensor Green's function at the particle horizon
$\chi=\pm1$:
\be
\bar h^{\rm (disc)}(\chi) = {A_0\ov2}\,\theta(1-|\chi|).
\lb{h_asym_a}
\ee
Vice versa, a Green's function discontinuity of this type implies
that $\f_0=0$.

In this Appendix we verify that the solution of the real space
equation\rf{sseq} for $\bar h(\chi)$ indeed has the form\rf{h_asym_a}.
We do not assume that $\Rnu$ is a small parameter.
To be specific, 
we consider the $\bar h(\chi)$ discontinuity at $\chi=-1$.

We take $\bar h(\chi)$ of the general form\rf{h_asym}
and first show that the right hand side of eq.\rf{sseq} should 
remain finite as $\chi\to-1+0$.
For the second term in the integral 
on the right hand side of eq.\rf{sseq},\footnote{
  Here is a proof of eq.\rf{finint2}.
  This equation considers 
  $\mbox{$(1-\chi^2)^2\int_{-1}^1d\mu\,\mu\,\bar h(\mu)/(\mu-\chi)$}$
  in the limit $\chi\to-1+0$. 
  The most divergent at $\chi\to-1$ contribution to 
  $\int_{-1}^1d\mu\,\mu\,\bar h(\mu)/(\mu-\chi)$
  arizes from the two discontinuous 
  $\bar h(\mu)$ terms\rf{h_asym}.
  The products of either of the two corresponding integrals,
  \be
  \qquad \int_{-1}^1d\mu\,\fr{\mu}{\mu-\chi}
   &=& \ln(\chi+1)\lf[1+o(1)\rt],
   \nn\\
  \int_{-1}^1d\mu\,\fr{\mu\ln|1-\mu^2|}{\mu-\chi}
    &=&  \fr12\ln^2(\chi+1)\lf[1+o(1)\rt],
  \nn
  \ee
  and of the prefactor $(1-\chi^2)^2$ in the considered expression
  apparently vanish as $\chi\to-1$.
}
\be
\lim_{\chi\to-1+0}\int_{-1}^1d\mu\,K(\chi,\mu) = 0.
\lb{finint2}
\ee
The remaining integral $\int_{-1}^1d\mu\,K(\mu,\chi)$ can be calculated 
explicitly.  However, even without its full calculation, 
it is easy to see that for $\chi\to-1$
\be
\int_{-1}^1d\mu\,K(\mu,\chi)= 
     -\fr43\,\bar h(\chi)\lf[1+o(1)\rt].
\ee
Thus we can write eq.\rf{sseq} as
\be
\lf[(\chi^2-1)\bar h'\rt]' = -2\Rnu\bar h(\chi)\lf[1+o(1)\rt].
\lb{heq_m1}
\ee

For $\chi>-1$, the last equation  is formally solved by
\be
\bar h = -2\Rnu 
   \int_{-1}^{\chi}\!\fr{d\chi_1}{\chi_1^2-1}
    \int_{-1}^{\chi_1}d\chi_2\,\bar h(\chi_2)\lf[1+o(1)\rt]+
\nn
\\
     +\ \const.\quad
\lb{barh_form_sol}
\ee
Given $\bar h$ of the general form\rf{h_asym},
the above double integral is {\it continuous\/} at $\chi=-1$.
Since the prefactor of~$\bar h''$ in eq.\rf{heq_m1}
vanishes at $\chi=-1$, the last constant in eq.\rf{barh_form_sol}
may differ for $\chi<-1$
and $\chi>-1$.  The change of the constant
provides a {\it finite\/} jump in $\bar h$ at $\chi=-1$.
This proves eq.\rf{h_asym_a}.

\subsection{Subleading in $1/k\t$ term}

The result\rf{barh_form_sol} allows to establish a simple
relation between the leading and subleading in $1/k\t$
terms that describe the subhorizon evolution of tensor 
modes during radiation
domination. Evaluating the 
right derivatives of both sides of this equation at $\chi=-1$,
we find a relation
\be
\bar h'(-1+0)=\Rnu\bar h(-1+0).
\ee 
Since $h(\chi)$ is even, we have a similar equality
at $\chi=1$: $\bar h'(1-0)=-\Rnu\bar h(1-0)$.
Remembering that $\bar h\equiv 0$ for $|\chi|>1$ 
and considering the Fourier components of $\bar h$,
we conclude that
\be
h^{\rm (rad)}(k\t)=
        A_0 \lf\{\fr{\sin k\t}{k\t}-\Rnu\fr{\cos k\t}{k^2\t^2}
        \lf[1+o(1)\rt]\rt\}.
\lb{h_rad_sublead_a}
\ee
This  relation applies to all orders of the $\Rnu$
expansion; its validity for $O(\Rnu)$ order
is evident from eqs.\rf{h_rad_fin} and\rf{h1k}.


\bibliography{tensbib}

\begin{thebibliography}{47}
\expandafter\ifx\csname natexlab\endcsname\relax\def\natexlab#1{#1}\fi
\expandafter\ifx\csname bibnamefont\endcsname\relax
  \def\bibnamefont#1{#1}\fi
\expandafter\ifx\csname bibfnamefont\endcsname\relax
  \def\bibfnamefont#1{#1}\fi
\expandafter\ifx\csname citenamefont\endcsname\relax
  \def\citenamefont#1{#1}\fi
\expandafter\ifx\csname url\endcsname\relax
  \def\url#1{\texttt{#1}}\fi
\expandafter\ifx\csname urlprefix\endcsname\relax\def\urlprefix{URL }\fi
\providecommand{\bibinfo}[2]{#2}
\providecommand{\eprint}[2][]{\url{#2}}

\bibitem[{\citenamefont{{Starobinskii}}(1979)}]{Starob79}
\bibinfo{author}{\bibfnamefont{A.~A.} \bibnamefont{{Starobinskii}}},
  \bibinfo{journal}{JETP Lett.} \textbf{\bibinfo{volume}{30}},
  \bibinfo{pages}{682} (\bibinfo{year}{1979}).

\bibitem[{\citenamefont{Rubakov et~al.}(1982)\citenamefont{Rubakov, Sazhin, and
  Veryaskin}}]{RubSazhVer82}
\bibinfo{author}{\bibfnamefont{V.~A.} \bibnamefont{Rubakov}},
  \bibinfo{author}{\bibfnamefont{M.~V.} \bibnamefont{Sazhin}},
  \bibnamefont{and} \bibinfo{author}{\bibfnamefont{A.~V.}
  \bibnamefont{Veryaskin}}, \bibinfo{journal}{Phys. Lett.}
  \textbf{\bibinfo{volume}{B115}}, \bibinfo{pages}{189} (\bibinfo{year}{1982}).

\bibitem[{\citenamefont{Fabbri and Pollock}(1983)}]{FabPol83}
\bibinfo{author}{\bibfnamefont{R.}~\bibnamefont{Fabbri}} \bibnamefont{and}
  \bibinfo{author}{\bibfnamefont{M.~d.} \bibnamefont{Pollock}},
  \bibinfo{journal}{Phys. Lett.} \textbf{\bibinfo{volume}{B125}},
  \bibinfo{pages}{445} (\bibinfo{year}{1983}).

\bibitem[{\citenamefont{Abbott and Wise}(1984)}]{AbbottWise84}
\bibinfo{author}{\bibfnamefont{L.~F.} \bibnamefont{Abbott}} \bibnamefont{and}
  \bibinfo{author}{\bibfnamefont{M.~B.} \bibnamefont{Wise}},
  \bibinfo{journal}{Nucl. Phys.} \textbf{\bibinfo{volume}{B244}},
  \bibinfo{pages}{541} (\bibinfo{year}{1984}).

\bibitem[{\citenamefont{Kamionkowski et~al.}(1997)\citenamefont{Kamionkowski,
  Kosowsky, and Stebbins}}]{Kamionk96_Bmobe}
\bibinfo{author}{\bibfnamefont{M.}~\bibnamefont{Kamionkowski}},
  \bibinfo{author}{\bibfnamefont{A.}~\bibnamefont{Kosowsky}}, \bibnamefont{and}
  \bibinfo{author}{\bibfnamefont{A.}~\bibnamefont{Stebbins}},
  \bibinfo{journal}{Phys. Rev. Lett.} \textbf{\bibinfo{volume}{78}},
  \bibinfo{pages}{2058} (\bibinfo{year}{1997}), \eprint{astro-ph/9609132}.

\bibitem[{\citenamefont{Seljak and Zaldarriaga}(1997)}]{SelZald_Bmode}
\bibinfo{author}{\bibfnamefont{U.}~\bibnamefont{Seljak}} \bibnamefont{and}
  \bibinfo{author}{\bibfnamefont{M.}~\bibnamefont{Zaldarriaga}},
  \bibinfo{journal}{Phys. Rev. Lett.} \textbf{\bibinfo{volume}{78}},
  \bibinfo{pages}{2054} (\bibinfo{year}{1997}), \eprint{astro-ph/9609169}.

\bibitem[{\citenamefont{Bardeen}(1980)}]{Bardeen80}
\bibinfo{author}{\bibfnamefont{J.~M.} \bibnamefont{Bardeen}},
  \bibinfo{journal}{Phys. Rev.} \textbf{\bibinfo{volume}{D22}},
  \bibinfo{pages}{1882} (\bibinfo{year}{1980}).

\bibitem[{\citenamefont{Kodama and Sasaki}(1984)}]{KS84}
\bibinfo{author}{\bibfnamefont{H.}~\bibnamefont{Kodama}} \bibnamefont{and}
  \bibinfo{author}{\bibfnamefont{M.}~\bibnamefont{Sasaki}},
  \bibinfo{journal}{Prog. Theor. Phys. Suppl.} \textbf{\bibinfo{volume}{78}},
  \bibinfo{pages}{1} (\bibinfo{year}{1984}).

\bibitem[{\citenamefont{Bertschinger}()}]{Bert93}
\bibinfo{author}{\bibfnamefont{E.}~\bibnamefont{Bertschinger}},
  \bibinfo{note}{in Proc.\ Les Houches School, Session LX, ed.\ R.~Schaeffer et
  al. (Netherlands: Elsevier 1996), astro-ph/9503125}.

\bibitem[{\citenamefont{Weinberg}(2004)}]{Weinb_nu}
\bibinfo{author}{\bibfnamefont{S.}~\bibnamefont{Weinberg}},
  \bibinfo{journal}{Phys. Rev.} \textbf{\bibinfo{volume}{D69}},
  \bibinfo{pages}{023503} (\bibinfo{year}{2004}), \eprint{astro-ph/0306304}.

\bibitem[{\citenamefont{Hawking}(1966)}]{Hawking66}
\bibinfo{author}{\bibfnamefont{S.~W.} \bibnamefont{Hawking}},
  \bibinfo{journal}{Astrophys. J.} \textbf{\bibinfo{volume}{145}},
  \bibinfo{pages}{544} (\bibinfo{year}{1966}).

\bibitem[{\citenamefont{{S.\ Bashinsky, U.\ Seljak}}(2004)}]{BS}
\bibinfo{author}{\bibnamefont{{S.\ Bashinsky, U.\ Seljak}}},
  \bibinfo{journal}{Phys. Rev.} \textbf{\bibinfo{volume}{D69}},
  \bibinfo{pages}{083002} (\bibinfo{year}{2004}).

\bibitem[{pro()}]{proof_adiab_causal}
\bibinfo{note}{See Appendix~B of Ref.\ct{BS} for a proof}.

\bibitem[{\citenamefont{Dicus et~al.}(1982)}]{Dicus_NuID}
\bibinfo{author}{\bibfnamefont{D.~A.} \bibnamefont{Dicus}}
  \bibnamefont{et~al.}, \bibinfo{journal}{Phys. Rev.}
  \textbf{\bibinfo{volume}{D26}}, \bibinfo{pages}{2694} (\bibinfo{year}{1982}).

\bibitem[{\citenamefont{Heckler}(1994)}]{Heckler_NuQED}
\bibinfo{author}{\bibfnamefont{A.~F.} \bibnamefont{Heckler}},
  \bibinfo{journal}{Phys. Rev.} \textbf{\bibinfo{volume}{D49}},
  \bibinfo{pages}{611} (\bibinfo{year}{1994}).

\bibitem[{\citenamefont{{Lopez} and {Turner}}(1999)}]{LopezTurner_NuQED}
\bibinfo{author}{\bibfnamefont{R.~E.} \bibnamefont{{Lopez}}} \bibnamefont{and}
  \bibinfo{author}{\bibfnamefont{M.~S.} \bibnamefont{{Turner}}},
  \bibinfo{journal}{\prd} \textbf{\bibinfo{volume}{59}},
  \bibinfo{pages}{103502} (\bibinfo{year}{1999}).

\bibitem[{\citenamefont{Seljak and Zaldarriaga}(1996)}]{cmbfast}
\bibinfo{author}{\bibfnamefont{U.}~\bibnamefont{Seljak}} \bibnamefont{and}
  \bibinfo{author}{\bibfnamefont{M.}~\bibnamefont{Zaldarriaga}},
  \bibinfo{journal}{Astrophys. J.} \textbf{\bibinfo{volume}{469}},
  \bibinfo{pages}{437} (\bibinfo{year}{1996}), \eprint{astro-ph/9603033}.

\bibitem[{\citenamefont{Turner et~al.}(1993)\citenamefont{Turner, White, and
  Lidsey}}]{TWL93}
\bibinfo{author}{\bibfnamefont{M.~S.} \bibnamefont{Turner}},
  \bibinfo{author}{\bibfnamefont{M.~J.} \bibnamefont{White}}, \bibnamefont{and}
  \bibinfo{author}{\bibfnamefont{J.~E.} \bibnamefont{Lidsey}},
  \bibinfo{journal}{Phys. Rev.} \textbf{\bibinfo{volume}{D48}},
  \bibinfo{pages}{4613} (\bibinfo{year}{1993}), \eprint{astro-ph/9306029}.

\bibitem[{\citenamefont{Ng and Speliotopoulos}(1995)}]{NgSpel94}
\bibinfo{author}{\bibfnamefont{K.-w.} \bibnamefont{Ng}} \bibnamefont{and}
  \bibinfo{author}{\bibfnamefont{A.~D.} \bibnamefont{Speliotopoulos}},
  \bibinfo{journal}{Phys. Rev.} \textbf{\bibinfo{volume}{D52}},
  \bibinfo{pages}{2112} (\bibinfo{year}{1995}), \eprint{astro-ph/9405043}.

\bibitem[{\citenamefont{Spergel et~al.}(2003)}]{WMAPSpergel}
\bibinfo{author}{\bibfnamefont{D.~N.} \bibnamefont{Spergel}}
  \bibnamefont{et~al.}, \bibinfo{journal}{Astrophys. J. Suppl.}
  \textbf{\bibinfo{volume}{148}}, \bibinfo{pages}{175} (\bibinfo{year}{2003}),
  \eprint{astro-ph/0302209}.

\bibitem[{\citenamefont{Wang}(1996)}]{Wang95}
\bibinfo{author}{\bibfnamefont{Y.}~\bibnamefont{Wang}}, \bibinfo{journal}{Phys.
  Rev.} \textbf{\bibinfo{volume}{D53}}, \bibinfo{pages}{639}
  (\bibinfo{year}{1996}), \eprint{astro-ph/9501116}.

\bibitem[{\citenamefont{Koranda and Allen}(1995)}]{KorAll94_h_analyt}
\bibinfo{author}{\bibfnamefont{S.}~\bibnamefont{Koranda}} \bibnamefont{and}
  \bibinfo{author}{\bibfnamefont{B.}~\bibnamefont{Allen}},
  \bibinfo{journal}{Phys. Rev.} \textbf{\bibinfo{volume}{D52}},
  \bibinfo{pages}{1902} (\bibinfo{year}{1995}), \eprint{astro-ph/9410049}.

\bibitem[{\citenamefont{Pritchard and Kamionkowski}(2004)}]{PritchKam04}
\bibinfo{author}{\bibfnamefont{J.~R.} \bibnamefont{Pritchard}}
  \bibnamefont{and}
  \bibinfo{author}{\bibfnamefont{M.}~\bibnamefont{Kamionkowski}}
  (\bibinfo{year}{2004}), \eprint{astro-ph/0412581}.

\bibitem[{\citenamefont{Grishchuk}(1975)}]{Grishch74}
\bibinfo{author}{\bibfnamefont{L.~P.} \bibnamefont{Grishchuk}},
  \bibinfo{journal}{Sov. Phys. JETP} \textbf{\bibinfo{volume}{40}},
  \bibinfo{pages}{409} (\bibinfo{year}{1975}).

\bibitem[{\citenamefont{{R.\ K.\ Sachs, A.\ M.\ Wolfe}}(1967)}]{SachsWolfe67}
\bibinfo{author}{\bibnamefont{{R.\ K.\ Sachs, A.\ M.\ Wolfe}}},
  \bibinfo{journal}{Astrophys. J.} \textbf{\bibinfo{volume}{147}},
  \bibinfo{pages}{73} (\bibinfo{year}{1967}).

\bibitem[{\citenamefont{{Doroshkevich}
  et~al.}(1977)\citenamefont{{Doroshkevich}, {Novikov}, and
  {Polnarev}}}]{DoroshNovikPoln77}
\bibinfo{author}{\bibfnamefont{A.~G.} \bibnamefont{{Doroshkevich}}},
  \bibinfo{author}{\bibfnamefont{I.~D.} \bibnamefont{{Novikov}}},
  \bibnamefont{and} \bibinfo{author}{\bibfnamefont{A.~G.}
  \bibnamefont{{Polnarev}}}, \bibinfo{journal}{Soviet Astronomy}
  \textbf{\bibinfo{volume}{21}}, \bibinfo{pages}{529} (\bibinfo{year}{1977}).

\bibitem[{\citenamefont{{Starobinskii}}(1985)}]{Starob84}
\bibinfo{author}{\bibfnamefont{A.~A.} \bibnamefont{{Starobinskii}}},
  \bibinfo{journal}{Soviet Astronomy Letters} \textbf{\bibinfo{volume}{11}},
  \bibinfo{pages}{133} (\bibinfo{year}{1985}).

\bibitem[{\citenamefont{{Polnarev}}(1985)}]{Poln85_tens_polar}
\bibinfo{author}{\bibfnamefont{A.~G.} \bibnamefont{{Polnarev}}},
  \bibinfo{journal}{Soviet Astronomy} \textbf{\bibinfo{volume}{29}},
  \bibinfo{pages}{607} (\bibinfo{year}{1985}).

\bibitem[{\citenamefont{Zaldarriaga and Seljak}(1997)}]{Zald_Selj_allsky97}
\bibinfo{author}{\bibfnamefont{M.}~\bibnamefont{Zaldarriaga}} \bibnamefont{and}
  \bibinfo{author}{\bibfnamefont{U.}~\bibnamefont{Seljak}},
  \bibinfo{journal}{Phys. Rev.} \textbf{\bibinfo{volume}{D55}},
  \bibinfo{pages}{1830} (\bibinfo{year}{1997}), \eprint{astro-ph/9609170}.

\bibitem[{\citenamefont{Hannestad}(2003)}]{Hannestad03_nu}
\bibinfo{author}{\bibfnamefont{S.}~\bibnamefont{Hannestad}},
  \bibinfo{journal}{JCAP} \textbf{\bibinfo{volume}{0305}}, \bibinfo{pages}{004}
  (\bibinfo{year}{2003}), \eprint{astro-ph/0303076}.

\bibitem[{\citenamefont{Allen et~al.}(2003)\citenamefont{Allen, Schmidt, and
  Bridle}}]{Allen03_mnu_signal}
\bibinfo{author}{\bibfnamefont{S.~W.} \bibnamefont{Allen}},
  \bibinfo{author}{\bibfnamefont{R.~W.} \bibnamefont{Schmidt}},
  \bibnamefont{and} \bibinfo{author}{\bibfnamefont{S.~L.}
  \bibnamefont{Bridle}}, \bibinfo{journal}{Mon. Not. Roy. Astron. Soc.}
  \textbf{\bibinfo{volume}{346}}, \bibinfo{pages}{593} (\bibinfo{year}{2003}),
  \eprint{astro-ph/0306386}.

\bibitem[{\citenamefont{Tegmark et~al.}(2004)}]{TegmarkSDSS03}
\bibinfo{author}{\bibfnamefont{M.}~\bibnamefont{Tegmark}} \bibnamefont{et~al.}
  (\bibinfo{collaboration}{SDSS}), \bibinfo{journal}{Phys. Rev.}
  \textbf{\bibinfo{volume}{D69}}, \bibinfo{pages}{103501}
  (\bibinfo{year}{2004}), \eprint{astro-ph/0310723}.

\bibitem[{\citenamefont{Barger et~al.}(2004)\citenamefont{Barger, Marfatia, and
  Tregre}}]{Barger03_mnu}
\bibinfo{author}{\bibfnamefont{V.}~\bibnamefont{Barger}},
  \bibinfo{author}{\bibfnamefont{D.}~\bibnamefont{Marfatia}}, \bibnamefont{and}
  \bibinfo{author}{\bibfnamefont{A.}~\bibnamefont{Tregre}},
  \bibinfo{journal}{Phys. Lett.} \textbf{\bibinfo{volume}{B595}},
  \bibinfo{pages}{55} (\bibinfo{year}{2004}), \eprint{hep-ph/0312065}.

\bibitem[{\citenamefont{Crotty et~al.}(2004)\citenamefont{Crotty, Lesgourgues,
  and Pastor}}]{Crotty04_mnu}
\bibinfo{author}{\bibfnamefont{P.}~\bibnamefont{Crotty}},
  \bibinfo{author}{\bibfnamefont{J.}~\bibnamefont{Lesgourgues}},
  \bibnamefont{and} \bibinfo{author}{\bibfnamefont{S.}~\bibnamefont{Pastor}},
  \bibinfo{journal}{Phys. Rev.} \textbf{\bibinfo{volume}{D69}},
  \bibinfo{pages}{123007} (\bibinfo{year}{2004}), \eprint{hep-ph/0402049}.

\bibitem[{\citenamefont{Seljak et~al.}(2004)}]{SeljakSDSS04}
\bibinfo{author}{\bibfnamefont{U.}~\bibnamefont{Seljak}} \bibnamefont{et~al.}
  (\bibinfo{year}{2004}), \eprint{astro-ph/0407372}.

\bibitem[{\citenamefont{Fogli et~al.}(2004)}]{Fogli04_mnu}
\bibinfo{author}{\bibfnamefont{G.~L.} \bibnamefont{Fogli}}
  \bibnamefont{et~al.}, \bibinfo{journal}{Phys. Rev.}
  \textbf{\bibinfo{volume}{D70}}, \bibinfo{pages}{113003}
  (\bibinfo{year}{2004}), \eprint{hep-ph/0408045}.

\bibitem[{\citenamefont{Bashinsky and Bertschinger}(2001)}]{BB_PRL}
\bibinfo{author}{\bibfnamefont{S.}~\bibnamefont{Bashinsky}} \bibnamefont{and}
  \bibinfo{author}{\bibfnamefont{E.}~\bibnamefont{Bertschinger}},
  \bibinfo{journal}{Phys. Rev. Lett.} \textbf{\bibinfo{volume}{87}},
  \bibinfo{pages}{081301} (\bibinfo{year}{2001}), \eprint{astro-ph/0012153}.

\bibitem[{\citenamefont{Bashinsky and Bertschinger}(2002)}]{BB}
\bibinfo{author}{\bibfnamefont{S.}~\bibnamefont{Bashinsky}} \bibnamefont{and}
  \bibinfo{author}{\bibfnamefont{E.}~\bibnamefont{Bertschinger}},
  \bibinfo{journal}{Phys. Rev.} \textbf{\bibinfo{volume}{D65}},
  \bibinfo{pages}{123008} (\bibinfo{year}{2002}).

\bibitem[{\citenamefont{Grishchuk and Sidorov}(1990)}]{Grishchuk_squeez90}
\bibinfo{author}{\bibfnamefont{L.~P.} \bibnamefont{Grishchuk}}
  \bibnamefont{and} \bibinfo{author}{\bibfnamefont{Y.~V.}
  \bibnamefont{Sidorov}}, \bibinfo{journal}{Phys. Rev.}
  \textbf{\bibinfo{volume}{D42}}, \bibinfo{pages}{3413} (\bibinfo{year}{1990}).

\bibitem[{\citenamefont{{Jones}}(1982)}]{Jones_GenFunc}
\bibinfo{author}{\bibfnamefont{D.~S.} \bibnamefont{{Jones}}},
  \emph{\bibinfo{title}{{The Theory of Generalized Functions}}}
  (\bibinfo{publisher}{Cambridge University Press}, \bibinfo{year}{1982}),
  \bibinfo{edition}{2nd} ed.

\bibitem[{sum()}]{summary_of_gen_trf}
\bibinfo{note}{The Fourier transforms of typical discontinuous functions are
  summarized in Table~II of\ct{BS}.}

\bibitem[{\citenamefont{Sahni}(1988)}]{Sahni88_h_analyt}
\bibinfo{author}{\bibfnamefont{V.}~\bibnamefont{Sahni}},
  \bibinfo{journal}{Class. Quant. Grav.} \textbf{\bibinfo{volume}{5}},
  \bibinfo{pages}{L113} (\bibinfo{year}{1988}).

\bibitem[{\citenamefont{Bond and Szalay}(1983)}]{BondSz83}
\bibinfo{author}{\bibfnamefont{J.~R.} \bibnamefont{Bond}} \bibnamefont{and}
  \bibinfo{author}{\bibfnamefont{A.~S.} \bibnamefont{Szalay}},
  \bibinfo{journal}{Astrophys. J.} \textbf{\bibinfo{volume}{274}},
  \bibinfo{pages}{443} (\bibinfo{year}{1983}).

\bibitem[{\citenamefont{{C.-P.\ Ma, E.~Bertschinger}}(1995)}]{MaBert95}
\bibinfo{author}{\bibnamefont{{C.-P.\ Ma, E.~Bertschinger}}},
  \bibinfo{journal}{Astrophys. J.} \textbf{\bibinfo{volume}{455}},
  \bibinfo{pages}{7} (\bibinfo{year}{1995}), \eprint{astro-ph/9506072}.

\bibitem[{\citenamefont{{Dodelson}}(2003)}]{Dod_book}
\bibinfo{author}{\bibfnamefont{S.}~\bibnamefont{{Dodelson}}},
  \emph{\bibinfo{title}{{Modern cosmology}}} (\bibinfo{publisher}{Academic
  Press}, \bibinfo{year}{2003}).

\bibitem[{\citenamefont{Bashinsky}(2004)}]{B04}
\bibinfo{author}{\bibfnamefont{S.}~\bibnamefont{Bashinsky}}
  (\bibinfo{year}{2004}), \eprint{astro-ph/0405157}.

\bibitem[{ver()}]{verify_suph_gauge_indep}
\bibinfo{note}{This can be verified using the explicit expressions for
  $\al^\mu$, \eg, from Appendix~A.3 of\ct{BS}.}



\end{thebibliography}

\end{document}